# Artificial Intelligence in Space

George Anthony Gal[1], Cristiana Santos[2], Lucien Rapp[3], Réka Markovich[2], Leendert van der Torre[2]

[1]Legal Parallax, LLC, USA
[2]University of Luxembourg
[3]Lucien Rapp, Université Toulouse Capitole 1, SIRIUS Chair

## 1 Introduction

Governance of space activities is being confronted with a progressive transformation associated with the emergence of satellite systems and space-based services utilizing AI, which includes ML. This chapter identifies and examines some of the fundamental legal challenges linked to using AI in the space domain. The legal challenges emanating from the reliance and use of AI in space necessitates ascertaining the existence of linkage between space systems and services using AI to a system of governing rules and guiding legal principles.

The nature of the space and satellite industry presents a quintessential use-case for AI. Essentially, virtually all space activities and ventures constitute fertile ground ripe for employing AI. Indeed, AI is ripe for use in Earth orbit activities like active debris removal ("ADR" ), near Earth ventures such as abiotic resource extraction, and deep space exploration. Generally, AI applications occur in two principal vectors, which are:

- *autonomous robots* (or space objects) – whether in the form of an autonomous spacecraft or a satellite constellation, autonomous or intelligent space objects possess the ability to not only collect, analyze, and use data for informational and operation purposes but they can also go where no human has gone or could go, collecting probes and data – to autonomous spacecraft and swarm intelligence, assisting space activities, such as mining and use of abiotic resources, exploration, in-orbit servicing (IoS), active debris removal (ADR), and protection of space assets which includes self-preservation from rogue and unknown natural space objects; and

- analyzing an, if necessary, acting upon *space big data*, related to (1) debris monitoring, ( 2) self-preservation based on potential threat by rogue and unknown natural objects in the space domain as well as perceived threats from another human manufactured object, (3)  predictive analytics of very-high resolution (VHR) satellite imagery,(4)  real-time geospatial data analysis ,and (5) an analysis of data products derived  from a convergence of a wide spectrum of sources (e.g. satellite, drone, IoT, UAVs imagery and location data). Space big data also enables space cloud computing services**,** in which a data is stored on space based assets[1.] Indeed, the development of AI-based technologies combined with space data can enhance the production, storage, access and diffusion of the data in outer space and on Earth.

Space is undergoing seismic shifts driven by New Space (promoting a Smart, Fast and Now Space)[2], the GAFA web giants, newcomers, venture capital firms and start-ups. There is significant growth in number of space activities, space objects, and space actors. However, **new challenges** emerge in the course of such active exploration and use, whilst deploying AI in space. Elaborating upon a forward-looking perspective, harnessing AI and ML technologies in accessing and exploring outer space, as well as engaging in space-enabled downstream commercial applications and services, will, in all likelihood, **span a broad array of intended and unintended consequences**. These consequences stem from the use and misuse of such technologies which cannot be downplayed or disregarded. Accordingly, the following quadrant risks that merits attention:

(i) privacy issues associated with the use of these technologies, e.g. citizen tracking and surveillance, potential re-identification of individuals, function creep, fake imagery, biased automatic decision-making and unjust discrimination based on nationality, gender, race, geographic localization, lack of transparency, etc.

(ii) liability issues emerging from potential damage created by autonomous spacecraft, e.g. collisions; or by hacking/malware that aims to weaponize AI, and consequences on space data (security of sensitive data stored in outer space, malicious data capture).

---

[1] To increase of data capacity, cost reduction of services and real-time access to data storage.
[2] https://www.satellitetoday.com/innovation/2019/07/03/moving-from-newspace-to-nowspace/

These risks are more acute when **acknowledging important facets in the space field**. Firstly, space is a service-and-needs-oriented market dominated mostly by demand and competitive **industry logics**, **without a main centralized regulatory body**. Secondly, there is an **increased ubiquitous repercussion of space activities on Earth**, since the benefits and solutions that space provides for the problems and needs of mankind are becoming ubiquitous[3] (transport, smart city management, security, agriculture, climate change monitoring, etc.). In this line, European Space Agency (ESA) estimates that for every Euro spent in the sector, there has been six Euros benefit to society. This correlation reflects a **more accentuated dependence of Earth from space-based services**.

The breadth of these space-based services – much of these AI-enabled – requires consideration of a broad range of legal and regulatory issues that the space industry alone cannot answer. Also, **UN Space Treaties** leave much uncertainty about permitted AI uses and activities in space. Clearly there is a need to develop or reinterpret 'rules of the road' to enable continued and legally compliant access for commercial and civilian actors in space.

The principal objectives of this Chapter include:
1. Identifying and discussing the potential risks and challenges associated with implemented AI and ML in space;
2. Analyzing to which extent the current *corpus iuris spatialis* (from the seventies) can still give answers to these risks and challenges and which methodology to follow onwards; and
3. Discussing how AI-based legal tools can support space law.

Consistent with these objectives, Section **2** examines the specificities of AI in space, describes distinct features from AI on Earth and demonstrate the usefulness and benefits of AI in space. **Section 3** analyzes certain legal, ethical and governance risks associated with AI in space. **Section 4** discusses limitations in the current space law legal framework relating to AI in space. **Section 5** offers a methodological approach for determining the legal regime applicable to AI in space while **Section 6** approaches legal AI-based tools that enable knowledge representation and reasoning at the service of space law**. Section 7** summarizes this analysis of AI in space.

## 2   Contextual dynamics of space and specificities of AI in Space

Space technology, data and services have become indispensable in the daily lives of Europeans as well as the majority of most global inhabitants. Space based services and activities also play an essential role in preserving the strategic and national security interests of many States. Europe seeks to cement its position as one of the major space faring powers by implementing extensive freedom of action in the space domain that encourages scientific and technical progress and support the competitiveness and innovation capacity of space sector industries.

---

[3] Hon. Philip E. Coyle, Senior Advisor, Center for Defense Information, http://spacesecurityindex.org/wp-content/uploads/2014/10/spacesecurityindexfactsheet.pdf

To boost the EU's space leadership beyond 2020, a Regulation[1] proposal establishes the space programme of the Union and the European Union Agency for the Space Programme. The proposed budget allocation of **16bn € for the post-2020 EU space** programme[2] has been received by the European space industry as a clear and strong signal of the political willingness to reinforce Europe's leadership, competitiveness, sustainability and autonomy in space.[3] AI is one area where Europe is exerting its leadership role in the space domain.

The use of AI in space capitalizes from the **contextual quadrant** called '**New Space***' which is creating a more complex and challenging environment at the physical, technological and operational realms. The current contextual dynamics of space and **specificities of space amenable to AI are discussed below.**

## 2.1 Contextual dynamics of space

Currently space is defined as *Space 4.0* which refers to this era of pro-activeness, open-mindedness to both technology disruption and opportunity[4] where trends include space big data (e.g. data imagery), predictive and geospatial analytics applied thereto. In particular, this era is backed up by AI-based technology, machine learning (ML), and Internet of Things (IoT). IoT is forecasted to be pervasive by 2025, with connected "things" driving a data explosion with sensors deployed by mega constellations of smallsats, (such as Hiber, Astrocast and Keplercars).

The use of such technologies promote a ***digital revolution***, unlocking access to space-based benefits:[5] the space industry is now moving toward leveraging full digitalization of *products* (high-performance spacecraft infrastructure; on board computers, antennas, microwave products), *new processes* (increasing production speed and decreasing failure rates; and *data uptake* (the ability to assess the data right away, distribution as well as for data analytics, processing, visualization and value adding, enabling Earth Observation (EO) to become part of the larger data and digital economy.

These space-based benefits (products-processes-data uptake) increase the ***repercussion of space activities on Earth***. A growing number of key economic sectors (in particular land and infrastructure monitoring, security, as well as the digital economy, transport, telecommunications, environment, agriculture and energy) use satellite navigation and EO systems.

***Space democratization and privatization*** reflect the access to and participation in space by space-faring nations and non-governmental entities such as privately owned juridical entities. Among space actors, the private **sector** currently accounts for 70% of space activity[6] (UNOOSA, 2018). This percentage will only increase given the emergence of new private actors who seek commercial opportunities in the exploration and exploitation of space and its resources thanks to frontier technologies, such as AI and the data revolution.[7]

---

[1] In a vote on 17 April 2019, the European Parliament endorsed a provisional agreement reached by co-legislators on the EU Space Programme for 2021-2027, bringing all existing and new space activities under the umbrella of a single programme to foster a strong and innovative space industry in Europe.https://data.consilium.europa.eu/doc/document/ST-7481-2019-INIT/en/pdf

[2] These benefits represent a return of investment for Europe of between 10 to 20 times the costs of the programme.

[3] This budget will be used first, to maintain and upgrade the existing infrastructures of Galileo and Copernicus, so that our systems remain at the top. Second we will adapt to new needs, such as fighting climate change, security or internet of things.

[4] ESA, What is Space 4.0? https://www.esa.int/About_Us/Ministerial_Council_2016/What_is_space_4.0, 2016 (accessed 4 of May 2019).

[5] https://spacenews.com/digital-endeavors-in-space/

[6] "(…) *Nowadays, private sector augments all segments of the space domain, from ground equipment and commercial space transportation to satellite manufacturing and Earth observation services*. UNOOSA, Inter-Agency Meeting on Outer Space Activities: 2018, Thirty-eighth session. http://www.unoosa.org/oosa/en/ourwork/un-space/iam/38th-session.html, 2018 (accessed 4 of May 2019).

New actors together with emerging new technologies such as AI develop *new global business models* driven by demand, such as satellite constellations), tourism, asteroid and lunar mining, *in-situ* resource utilization[8] (ISRU), 5G, in-orbit servicing (IoS), 3D printing of satellite parts (e.g. solar panels, etc.),and commercial space station. These new business segments[9] are leveraging space economy. The space economy is expanding enormously, with predictions that it generates revenues of US$ 1.1-2.7 trillion or more by 2040.[10].

*New high-end technologies embedding small-satellite design* describe the current landscape of the space industry. Smaller, lightweight satellites based on affordable off-the-shelf hardware, less expensive spacecraft (small, nano and pico-satellites) can be replaced more easily thereby refreshing technology rapidly[11] combined with the ability to launch thousands of these satellites into mega constellations opens up possibilities for more missions and applications using space infrastructure.

### 2.2 Specificities of space amenable to AI

It is still important to consider the *specificities* of AI in outer space and why it is distinct from the its terrestrial use. Some of the amenable specificities and distinctions are as follows:

i. *Space conditions are hard and amenable only for AI machines*. Space is a remote, hostile and hazardous environment[12] to human life and in some cases impossible for humans to explore and survive in space renders space technologies dependent on technology and processes related to AI[13]. AI-based technologies fit for operational decision-making, which are robust, resilient, adaptable and responsive to changing threats.

ii. *Upstream and downstream impact of AI in space.* AI in a fast-approaching future will impact all sectors of the space industry, from launch to constellation control and satellite performance analysis[14], from AI logic directly on board the payload for deep space applications ranging to the downstream sector of telecommunications and Earth observation in commercial applications, e.g. for image classification and preditive analysis of phenomena.

iii. *Autonomy of intelligent space objects*. Using AI, a spacecraft may be able[15] to recognize a threat, capture data, learn from and counteract with it or take evasive action, and even propagate its newly acquired knowledge to other satellites. For example, ''*when a Mars rover conducting exploration of Mars needs to contact Earth, it takes up to 24 minutes to pass the signal between the two planets in one direction. It is rather long time for making decisions, that is why engineers are increasingly providing space robots with*

---

[7] European Investment Bank, The future of the European space sector How to leverage Europe's technological leadership and boost investments for space ventures, https://www.eib.org/attachments/thematic/future_of_european_space_sector_en.pdf

[8] Lucas-Rhimbassen M., Santos C., Long G., Rapp L., Conceptual model for a profitable return on investmentfrom space debris as abiotic space resource , 8TH European Coference for Aeronautics and Space Sciences (EUCASS), 2019.

9 And others, like Scalability and agility, Media / Advertising, B2C , Vertical integration, Position in value chain.

[10](http://www.unoosa.org/res/oosadoc/data/documents/2019/stspace/stspace76_0_html/UNOOSA_Annual_Report_2018.pdf

[11]Livemint, Mini satellites, maximum possibilities. https://www.livemint.com/Leisure/yEXAKO6k0UWRLtV6rzdQaP/Mini-satellites-maximum-possibilities.html, 2018 (accessed 4 of May 2019).

[12] E.g. difficult accessibility, the complexity of extra-atmospheric missions, the extreme physical and climatic conditions, new gravitational forces, different temperature ranges and unknown collisions with dust or an asteroid.

[13] Larysa Soroka, Kseniia Kurkova, Artificial Intelligence and Space Technologies: Legal, Ethical and Technological Issues, Advanced Space Law, Volume 3, 2019: 131-139.

[14] http://interactive.satellitetoday.com/via/december-2019/space-2-0-taking-ai-far-out/

[15] https://spacenews.com/self-driving-spacecraft-the-challenge-of-verifying-ai-will-work-as-intended/

*the ability to make decisions by themselves.[16]* AI provides space objects with the ability to collect and analyze data and decide when and what information to send back to Earth without any human involvement, and to predict, self-diagnose problems, and fix themselves while continuing to perform[17]. When collisions occur between intelligent space objects and debris, a legal issues like liability are triggered, some of which are dealt with in Section 3.1 and 4.

iv. **Asset Protection.** AI assists with the protection of space assets by allowing the development of automatic collision avoidance system that will assess the risk and likelihood of in-space collisions, improving the decision making process on whether or not an orbital manoeuvre is needed, and transmitting warnings to other potential at risk space objects.[18]

v. *Big data from Space*. Big data from space[19] refers to the massive spatio-temporal Earth and Space observation data collected by a variety of sensors – ranging from ground based to space-borne – and the synergy with data coming from other sources and communities. Spatial big data, when combined with "*Big Data Analytics,*" delivers "value" from big datasets, whose volume, velocity, variety, veracity, and value is beyond the ability of traditional tools to capture, store, manage and analyse the sheer volume of data. *Geospatial intelligence* is one of many ways to use artificial intelligence in outer space. It refers to employing AI for extracting and analyzing images and other geospatial information relating to terrestrial, aerial, and/or spatial objects and events. It also allows for real time interpretation of what is happening or transpiring in a specific geolocation relating to events such as disasters, refugee migration and safety, and agricultural production. These aspects are analysed in Section 3.2 of this Chapter.

# 3 Risks of AI in Space

AI in space is igniting a gradual shift from "computer-assisted human choice and human-ratified computer choice"[20] to non-human analysis, decision-making and implementation of action. The emerging deployment and use of intelligent space objects[21] present novel challenges to the current space law regime especially when, not if, the use of such objects causes terrestrial and/or extraterrestrial injury by an AI system or service such as a violation of privacy rights, violation of data protections requirement, or injury resulting from al collision involving a space object.[22]

---

[16] Downer, Bethany. The Role of Artificial Intelligence in Space Exploration, 2018. https://www. reachingspacescience.com/single-post/AI-in-SpaceExploration

[17] http://interactive.satellitetoday.com/via/december-2019/space-2-0-taking-ai-far-out/

[18] https://www.esa.int/Safety_Security/Space_Debris/Automating_collision_avoidance

[19] P. Soille, S. Loekken, and S. Albani (Eds.) Proc. of the 2019 conference on Big Data from Space (BiDS'2019), EUR 29660 EN, Publications Office of the European Union, Luxembourg, 2019, https://www.bigdatafromspace2019.org/QuickEventWebsitePortal/2019-conference-on-big-data-from-space-bids19/bids-2019

[20] Mariano-Florentino Cuellar, *A Simpler World? On Pruning Risks and Harvesting Fruits in an Orchard of Whispering Algorithms*, 51 U.C. Davis Law Review, 27, 39 (Nov. 2017).

[21] A space object is limited to an object, including its component parts which were "launched" into space).The issue can become a bit murkier if intelligent space objects can be manufactured and deployed in situ in outer space.

[22] It would be naive to think that intelligent space objects will not cause any injury. The experience associated with implementing autonomous motor vehicles should dispel any such notion. Emily Stewart, *Self-driving cars have to be safer than regular cars. The question is how much* vox.com (May 17, 2019) https://www.vox.com/recode/2019/5/17/18564501/self-driving-car-morals-safety-tesla-waymo

The space law treaty regime consists of the foundational Treaty on Principles Governing the Activities of States in the Exploration and Use of Outer Space, including the Moon and Other Celestial Bodies ("Outer Space Treaty")[23] and its progeny treaties. The OST embeds the cornerstone principles for the current international space law jurisprudence.[24] Its principles are expanded by the progeny treaties of the Agreement on the Rescue of Astronauts, the Return of Astronauts and the Return of Objects Launched into Outer Space ("Rescue Agreement"),[25] the Convention on International Liability for Damage Caused by Space Objects ("Liability Convention"),[26] the Convention on Registration of Objects Launched into Outer Space ("Registration Convention"),[27] and the Agreement Governing the Activities of States on the Moon and Other Celestial Bodies ("Moon Treaty").[28] Liability issues associated with AI risks concern an analysis of the Outer Space Treaty and the Liability Convention.

## 3.1 Liability of Intelligent Space Objects

Liability under the space law treaty regime is rooted in Outer Space Treaty Article VIII which is the genesis of the Liability Convention. Article VII imposes international liability only on a launching State [29] The Liability Convention establishes a restricted framework for assessing international liability which only applies to a launching States.[30] The determination of liability and allocation of faulty is based on where the damage occurs. Liability Convention Article II imposes absolute or strict liability for damage a space object causes on Earth or to an aircraft in flight. On the other hand, if a space object causes damage in outer space or on a celestial body, then liability is based on the degree of fault as allocated by Article III. This section applied these rules on liability in the context of intelligent space objects.

### 3.1.1 Some notes on liability Associated with Intelligent Space Objects

The "damage" covered by the Liability Convention is neither comprehensive nor unambiguous. Article 1(a) defines "damage" to mean "**loss of life, personal injury or other impairment of health; or loss of or damage to property of States or of persons, natural or juridical, or property of international intergovernmental organizations.**" This definition creates uncertainty regarding the parameters or scope of damage subject to the Convention. It is unclear if the damage is limited to physical damage caused by space object[31] and if it extends to non-kinetic harm, indirect damage or purely economic injury.[32] Similarly, it is unsettled how far the phase "other impairment of health" reaches in connection with a person. For instance, is the phrase limited to physical injury or if it extends to emotional and/or mental injury?

---

[23] entered into Force Oct. 10, 1967, 18 UST 2410; TIAS 6347; 610 UNTS 205; 6 ILM 386 (1967).
[24] Frans von der Dunk, *Sovereignty Versus Space - Public Law and Private Launch in the Asian Context*, 5 Singapore Journal of International and Comparative Law 22, 27 (2001).
[25] entered into Force Dec. 3, 1968, 19 UST 7570; TIAS 6599; 672 UNTS 119; 7 ILM 151 (1968)
[26] entered into Force Sept. 1, 1972, 24 UST 2389; TIAS 7762; (961 UNTS 187; 10 ILM 965 (1971)
[27] entered into Force Sept. 15, 1976, 28 UST 695; TIAS 8480; 1023 UNTS 15; 14 ILM 43 (1975)
[28] entered into Force July 1, 1984, 1363 UNTS 3; 18 ILM 1434 (1979). The Moon Treaty is viewed differently than the other space treaties because it has not received the international ratification of the other space law treaties and the major space faring nations such as the United States, Russia and China have neither signed nor ratified the treaty
[29] Liability Convention Article 1( c) defines the term **"launching State"** as a State which launches or procures the launch of the space object and the State from whose territory or facility the space object is launched. A non-governmental space actor does not have international liability under the Liability Convention for damage caused by the space object regardless of their culpability. This means that a State space actor can only have international liability if it comes within the definition of a "launching State."
[30] George Anthony Long, *Artificial Intelligence and State Responsibility Under the Outer Space Treaty* at 5, 69th IAC, Bremen, Germany, ( Oct 5, 2018) published in 2018 Proceedings Of The International Institute Of Space Law (Eleven Int'l Publishing 2018).
[31] Major Elizabeth Seebode Waldrop, *Integration of Military and Civilian Space Assets: Legal and National Security Implications*, 55 A.F.L. Rev. 157, 214 (2004).
[32] George Anthony Long, Small Satellites and State Responsibility Associated With Space Traffic Situational Awareness at 3, 1st Annual Space Traffic Management Conference "Roadmap to the Stars," Embry-Riddle Aeronautical University, Daytona Beach, Fla., Nov. 6, 2014 available at https://commons.erau.edu/stm/2014/thursday/17

Resolution of the reach of a damage claim, like all legal issues associated with the Liability Convention, depends upon whether the definition is given a restrictive or expansive interpretation. Intelligent space objects, i.e., autonomous space objects utilizing AI, present challenges for the strict and fault liability scheme imposed on launching States.

Liability Convention Article III reads as follows:

> [i]n the event of damage being caused elsewhere than on the surface of the Earth to a space object of one launching State or to persons or property on board such a space object by a space object of another launching State, **the latter shall be liable only if the damage is due to its fault or the fault of persons** for whom it is responsible. (Emphasis added)

Intelligent space objects disrupt Article III's fault-based liability scheme since decisions, acts or omissions of an intelligent space object may not be construed to be conduct of a person and may not always be attributable to a launching State.

### 3.1.2 Fault Liability is Predicated on Human Fault

Generally, we think of a person as a human being.[33] In the legal arena, the term "person" generally refers to an entity which is subject to legal rights and duties.[34] The law considers artificial entities like corporations, partnerships, joint ventures, and trusts to be a "person" as they are subject to legal rights and duties.[35] Additionally, in certain instances, the law recognizes and imposes legal rights and duties on certain inanimate objects like ships, land, and goods which results in such inanimate objects being subject to judicial jurisdiction as well as being subject to a judgment rendered against it.[36] However, the legal rights and duties imposed on artificial entities and inanimate objects flow from actions or conduct engaged in by human beings.

This is not necessarily the case for actions or conduct taken based on **machine intelligence**. Although a machine can learn independently from human input and make decisions based on its learning and available information, that ability does not necessarily equate with natural or legal personhood. As noted, decisions and conduct of legal persons are ultimately decisions made by a human being. This means the decision is not based solely on intellect or data but is also the product of human factors such as a conscious, emotion, and discretion.[37] Thus, the concept of legal personhood is ultimately premised on humanity. Decisions and conduct based on AI divorced from human oversight or control arguably lack consideration of human factors such as a conscious, emotion and discretion.[38] Even more so, there is not currently any law which grants "personhood" to an intelligent space object. The lack of direct or indirect human considerations in the decision making of an intelligent machine together with such an object not having any legal rights or duties under existing law strongly suggests that decisions by an intelligent space object are not made by a natural or legal person.[39]

Since fault liability under Liability Convention Article III is premised on the fault of a State or the faults of persons, a decision by an intelligent space object will, in all likelihood, not be the "fault of persons."

---

[33] Lawrence B. Solum, *Legal Personhood for Artificial Intelligences*, 70 N.C.L. Rev. 1231, 1238 (1992).
[34] *Id*., at 1238-1239.
[35] *Id*.
[36] *Id*., at 1239.
[37] *Id*.,at 1262 - 1287.
[38] *Id*.
[39] Long, *supra* Note 39, at 7.

Accordingly, assessing fault liability under Article III for a decision made by an intelligent space object may very well depend upon whether such a decision can be attributable to the launching State.

### 3.1.3 Fault Liability in Absence of Human Oversight in the Decision Making Process

Generally, liability for damage or injury attributable to States is traceable to human acts or omissions. This basis for imposing liability appears to be inapplicable when damage or injury in outer space is caused by an analysis, decision, and implementation of a course of action made by a machine without human approval.[40]

Liability premised on human acts or omission fails when no particular human possessed the ability to prevent the injury, short of making the decision to utilize AI in a space object.[41] For sure, it is substantively difficult to draw a line between reliance on AI to supplant the judgment of a human decision maker and the propriety of allowing a machine, or nonhuman, to decide and implement a course of action.[42] To this extent, it seems that launching State fault-based liability should not be premised solely on the decision to launch an intelligent space object, as such a sweeping basis for liability would effectively retard the development and deployment of intelligent space objects.[43] Thus, the appropriate analysis appears to be what conduct is necessary to attribute fault liability to a State for damage caused by an intelligent space object when human oversight is not involved in the occurrence causing the damage.

Resolving this dilemma presents novel and complex issues associated with the standard of care, foreseeability and proximate cause which are crucial elements for establishing fault (under Liability Convention Article III).[44] This matter is further complicated by the distinct possibility that it may not be possible to ascertain how an intelligent space object made a particular decision.[45]

Nevertheless, untangling these nuanced legal obstacles may not be necessary to assess fault liability. Outer Space Treaty Article VI requires a State to assure that the space activities of its governmental and non-governmental entities comply with the Outer Space Treaty. It not only makes a State internationally responsible for its national activities in outer space, but it also imposes a dual mandate of "authorization and continuing supervision" that is not limited to the launching State or the space actor's home State.[46]

Outer Space Treaty Article VI does not expressly burden the launching State with the authorization and supervision obligation. Instead, it vests the authorization and continuous supervision on the "**appropriate State.**" Since neither Outer Space Treaty Article VI nor any other provision of the space law treaty regime define the term "appropriate State", or sets forth any criteria for ascertaining the appropriate State(s), there are not any agreed upon legal standards for determining what constitutes the "appropriate State". Nevertheless, it has been articulated that **a launching State is generally always an appropriate party** for purposes of Outer Space Treaty Article VI.[47] This is a reasonable and accurate extrapolation since the liability scheme is predicated on launching State status.

---

[40] Curtis E.A. Karnow, *Liability For Distributed Artificial Intelligences*, 11 Berkeley Technology Law Journal 147, 189-190 (1996).
[41] *Id*.
[42] *Id*.
[43] *See Long, supra* Note 39, at 7. *See also* Weston Kowert, *The Foreseeability of Human-artificial Intelligence Interactions,* 96 Texas Law Review 181, 183 (2017).
[44] *Long, supra* Note 39, at 8. While the decision to launch an intelligent space object may not be the basis for fault liability, as discussed infra, how the decision was made may serve as a vehicle for assessing fault liability
[45] *Id*.
[46] *See Id.*
[47] *See Generally* Bin Cheng, *Article VI Of The 1967 Space Treaty Revisited: "International Responsibility," "National Activities," And The Appropriate State."* 26 Journal of Space Law 7 (1998)

Since fault liability is generally predicated on the breach of a standard of care,[48] the dual responsibility of "authorization and continuing supervision by the appropriate State party" arguably establishes a standard of care which a launching State must comply with,[49] especially in connection with an intelligent space object. This essentially means that **a launching State bears the duty to ensure that appropriate authorization and supervision will be exercised in connection with an intelligent space object that it launches for a non-governmental entity, regardless of whether the object is owned or operated by a national**. The standard of care analysis, therefore, shifts from the specific occurrence which caused the damage to examine whether the launching State exercised sufficient authorization and supervision over the activities engaged in by the intelligent space object.

By analogizing to the **"due diligence"** standard under international law,[50] determining whether a launching State exercised sufficient authorization and supervision involves a **flexible** and fluid standard. "Due diligence" is not an obligation to achieve a particular result; rather it is an obligation of conduct which requires a State to engage in sufficient efforts to prevent harm or injury to another State or its nationals[51] or the global commons.[52] The breach of this duty is not limited to State action, but it also extends to the conduct of a State's nationals.[53] While there is "an overall minimal level of vigilance" associated with due diligence, "a higher degree of care may be more realistically expected" from States possessing the ability and resources to provide it.[54] In any event it would appear that a launching State's standard of care entails assuring that there is some State authorization and supervision over the space activities engaged in by the intelligent space object. However, based on the **flexible** standard of care, it seems that the function of the intelligent space object determines whether human input or oversight is required and if so, what is the appropriate degree and extent of the human oversight.

This flexibility is consistent with the approach recognized by the **European Commission** (EC) in connection with Artificial Intelligence in general.[55] The EC has adopted the policy **that human oversight is a necessary component in the use of AI**.[56] The policy is premised on the reasoning that human oversight ensures that an "AI system does not undermine human autonomy or cause other adverse effects."[57] The human oversight requires the "appropriate involvement by human beings" which may vary depending upon the "intended use of the AI system" and the "effect," if any, it may have on people and legal entities.[58] The EC then enumerates certain non-exhaustive manifestations of human oversight which include 1) human review and validation of an AI decision either before or immediately afterward implementation of the decision, 2) monitoring of the AI system "while in operation and the ability to intervene in real time and deactivate" the AI system, and 3) imposing operational restraints to ensure that certain decisions are not made by the AI system.[59] This EC policy presents a flexible framework that can be utilized to determine whether a launching State has met its standard of care in relation to a non-governmental intelligent space object that causes damage in outer space.

---

[48] Joel A. Dennerley, *supra Note 3*, 29 Eur. J. Intl. L. at 295.
[49] *See Generally Bin Cheng, supra* Note 67.
[50] Dennerley, *supra* Note 3, at 293-295.
[51] ILA Study Group *supra* Note 71 at 29 citing Responsibilities and Obligations of States Sponsoring Persons and Entities with Respect to Activities in the Area, Seabed Mining Advisory Opinion at ¶ 117 (Seabed Dispute Chamber of the International Tribunal of the Law of the Sea, Case No 17 , 1 February 2011); Jan E. Messerschmidt, *Hackback: Permitting Retaliatory Hacking by Non-State Actors As Proportionate Countermeasures to Transboundary Cyberharm Shearman & Sterling Student Writing Prize in Comparative and International Law,* Outstanding Note Aw, 52 Colum. J. Transnat'l L. 275, 302 - 305 (2013). *See United States Diplomatic and Consular Staff in Tehran (U.S. v. Iran)*, 1980 I.C.J. 3, 61 - 67 (May 24).
[52] *See* Mark Allan Gray, *The International Crime of Ecocide*, 26 Cal. W. Int'l L.J. 215, 238 (1996). at 242; Robert Rosenstock and Margo Kaplan*, The Fifty-Third Session of the International Law Commission*, 96 Am. J. Int'l L. 412, 416 (2002).
[53] Mark Allan Gray, *supra* Note 73, 26 Cal. W. Int'l L.J. at 243.
[54] *Id*.; *See* ILA Study Group *supra* Note 71 at 4 and 31.
[55] *European Commission, White Paper on Artificial Intelligence - A European approach to excellence and trust*, COM(2020) 65 final (Brussels, 19.2.2020) available at
https://ec.europa.eu/info/sites/info/files/commission-white-paper-artificial-intelligence-feb2020_en.pdf
[56] *Id*., at 21.
[57] *Id*.
[58] *Id*.

The due diligence flexible standard can also be used by the launching State to negate or mitigate its liability for damage caused by an intelligent space object. The flexible standard will allow a launching State to argue that the home State of the non-governmental space actor bears a greater degree of oversight responsibility than the launching State. Accordingly, it should be reasonable and sufficient for a launching State to rely on assurances that the non-national's home State exercises adequate authorization and oversight procedures for its nationals' use of intelligent space objects. This shifts the supervisory obligation from the launching State to the home State of the non-governmental space actor. The home State's failure to properly exercise its standard of care may, depending upon the circumstance, mitigate or absorb the launching State of fault liability under Liability Convention Article III. This shift, however, is not automatic as the due diligence standard makes it dependent on the home State's technological prowess in the area of AI or its financial ability to contract for such expertise.

### 3.1.4 Intelligent Space Objects and Absolute Liability

Liability Convention Article II imposes strict liability on a launching State if a space object causes damage on the Earth's surface or to aircraft in flight. Article VI(1), however, allows **exoneration** from absolute liability if the damage results "either wholly or partially from gross negligence or from an act or omission done with intent to cause damage on the part of a claimant State or of natural or juridical persons it represents." This defense, however, may not be available if the damage results "wholly or partially" from an act or omission of an intelligent space object deployed or controlled by the claimant State or a natural or juridical person the claimant State represents.

"Gross negligence" the mental element of an act or omission are products of human thought, which are absent in the machine decision process.[60] Even more so, Liability Convention Article VI may also defeat exoneration from absolute liability if the claimant State is able to show that the launching State's deployment of the intelligent space object breached its State responsibility under international law, the United Nations Charter or the Outer Space Treaty. This defense to the negation of absolute liability thrusts consideration of Outer Space Treaty Article VI into the equation.

### 3.1.5 Intelligent Space Objects and Liability Under Outer Space Treaty Article VII

Outer Space Treaty Article VII imposes international liability on the launching State, without qualification or exception. Moreover, Article VII does not predicate fault liability on a human involvement in the damage causing event or fault being otherwise attributable to the launching State. This unqualified launching State liability may possibly present an alternative recourse for pursuing compensation for damage in space caused by an intelligent space object. The pursuit of monetary compensation under Outer Space Treaty Article VII may very well arise when fault cannot be assessed under Liability Convention Article III because the decision causing the damage was not made by a person and the decision is not otherwise attributable to a launching State. The issue can also surface if a claimant State seeks financial compensation for an injury or harm cause by an intelligent space object which does not come within the meaning of "damage" as defined by Liability Convention Article 1(a). For instance, if an intelligent space object is used to interfere with, jam or hijack a commercial satellite transmission, then the financial injury suffered as a consequence of such conduct may not be compensable under the Liability Convention given its definition of "damage." Outer Space Treaty Article VII, however, may provide a basis for recovery under such a circumstance.

---

[59] *Id*.
[60] *See supra* at 9-10.

Of course, a party seeking to pursue such a remedy under Outer Space Treaty Article VII may, in all likelihood, encounter the objection that since the Liability Convention is the progeny of Outer Space Treaty Article VII, a State is estopped from pursuing a remedy directly under Outer Space Treaty Article VII. Such an objection can be premised on the public international law principle that "when a general and a more specific provision both apply at the same time, preference must be given to the special provision."[61] It is unclear if this principle can apply to the relationship between Outer Space Treaty Article VII and the Liability Convention.

Although the Liability Convention expressly proclaims that one of its principal purposes is to establish rules and procedures "concerning liability for damage caused by space objects,"[62] the treaty does not assert that its rules and procedures are exclusive when assessing liability through means other than the Liability Convention.. Most important though, is that neither the Outer Space Treaty nor the Liability Convention precludes a recovery of damage under Outer Space Treaty Article VII. This point is significant given the general principle of international law that what is not prohibited is permitted.[63] In other words, "'in relation to a specific act, it is not necessary to demonstrate a permissive rule so long as there is no prohibition.'"[64]

The determination of whether Liability Convention Article III estops a State from pursuing recourse under Outer Space Treaty Article VII for an injury caused by a space activity is, like most current space law issues, purely an academic exercise in as much as there is scant guidance from national or international courts, tribunals, or agencies on interpreting the provisions of the space law treaty regime. Nevertheless, resolution of the issue presents a binary choice of whether the Liability Convention does or does not preclude resort to Outer Space Treaty Article VII. The resolution of the issue will have a significant impact on whether the Liability Convention needs to be amended or supplemented to accommodate the deployment and use of intelligent space objects. For sure, if relief can be obtained under Outer Space Treaty Article VII when a remedy is not available under the Liability Convention, then Outer Space Treaty Article VII should provide sufficient flexibility to address liability issues associated with intelligent space objects during this period of AI infancy.

### 3.2 Data protection and ethical challenges related to AI in Space

Every year, commercially available satellite images are becoming **sharper** and taken more frequently. The leading-edge imagery resolution commercially available limits each pixel in a captured image to approximately 31 cm[65]. There is increasing demand from private commercial entities pushing[66] for lowering the resolution restrictions threshold to 10 cm[67]-[68]. The significance of using AI in connection with satellite imaging is best illustrated by the United States, in January 20202, imposing immediate interim export controls regulating the dissemination of AI technology software that possesses the ability to automatically scan aerial images to recognize anomalies or identify objects of interest , such as vehicles, houses, and other structures.[69]

---

[61] Alessandra Perrazzelli and Paolo R. Vergano, *Terminal Dues Under the Upu Convention and the Gats: An Overview of the Rules and of Their Compatibility*, 23 Fordham Intl. L.J. 736, 747 (2000)
[62] Liability Convention Preamble, 4th Paragraph. The other purpose is to ensure prompt payment "of a full and equitable measure of compensation to victims" in accordance with the Convention.
[63] *S.S. Lotus*, P.C.I.J. Ser. A, No. 10 at 18 (1927).
[64] Roland Tricot and Barrie Sander, "*Recent Developments: The Broader Consequences Of The International Court of Justice's Advisory Opinion On The Unilateral Declaration of Independence In Respect Of Kosovo,*" 49 Columbia Journal of Transnational Law, 321, 327 (2011) quoting Accordance with International Law of the Unilateral Declaration of Independence in Respect of Kosovo (Kosovo Advisory Opinion), Advisory Opinion, 2010 I.C.J. at 2 (July 22)(declaration of Judge Simma).
[65] http://worldview3.digitalglobe.com/
[66] https://www.nextbigfuture.com/2019/09/us-spy-satellites-at-diffraction-limit-for-resolution-since-1971.html
[67] https://www.nextbigfuture.com/2019/09/us-spy-satellites-at-diffraction-limit-for-resolution-since-1971.html
[68] https://www.bbc.com/future/article/20140211-inside-the-google-earth-sat-lab?referer=https%3A%2F%2Fwww.businessinsider.com%2Fsatellite-image-resolution-keeps-improving-2015-10%3Finternational%3Dtrue%26r%3DUS%26IR%3DT

**Speculation** revolves around satellite imagery discerning car plates, individuals, and "manholes and mailboxes"[70]. In fact, in 2013, police in Oregon, used Google Earth satellite image depicting marijuana growing illegally on a man's property[71]. In 2018, Brazilian police used real-time satellite imagery[72] to detect a spot where trees had been ripped out of the ground to illegally produce charcoal and arrested eight people in connection with the scheme. In China, human rights activists used satellite imagery[73] to show that many of the Uighur reeducation camps in Xinjiang province are surrounded by watchtowers and razor wire. A recent case deployed ML to create a system that could autonomously review video footage and detect patterns of activity and in one of the test cases, the system monitored video of a parking lot and identified moving cars and pedestrians. This system established a baseline of normal activity from which anomalous and suspicious actions could be detected.[74]

Even if this image and video resolution do not suffice to be able to distinguish[75] individuals or their features, it is **no longer in a sweet spot**. The broad **definition of personal data** included in the General Data Protection Regulation[76] (GDPR), enables that *all* information of EO data related to an identified or identifiable natural person (as location data) can be considered as personal data[77]. The broad **definition of personal data** included in the GDPR enables that *all* information of EO data related to an identified or identifiable natural person (as location data) can be considered as personal data[78]. The attribute "*identified*" refers to a known person, and "*identifiable*" refers to a person who is not identified yet, but identification is still possible. An individual is directly identified or identifiable by reference to "*direct or unique identifiers*". These "direct and unique identifiers" covers data types which can be easily referenced and associated with an individual, including descriptors such as a name, an identification number or username, location data, card of phone numbers, online identifiers, etc., as described in the GDPR. An individual is "*indirectly identifiable*" by combinations of indirect (and therefore non-unique identifiers) that allow individual to be singled out; they are less obvious information types which can be related to, or "linked" to an individual, such as, for instance, video footage, public key, signatures, IP addresses, device identifiers, metadata, and alike.

---

[69] 85 Fed. Reg. 459 (January 6, 2020)
[70] See US lifts restrictions on more detailed satellite images, BBC, http://www.bbc.com/news/technology-27868703
[71] https://www.cbsnews.com/news/google-earth-used-to-bust-oregon-medicinal-marijuana-garden-police-say/
[72] https://blog.globalforestwatch.org/people/amapa-police-use-forest-watcher-to-defend-the-brazilian-amazon
[73] https://www.reuters.com/investigates/special-report/muslims-camps-china/
[74] See https://aerospace.org/Annual-Report-2018/artificial-intelligence-gets-ahead-threats
[75] Cristiana Santos, Delphine Miramont, Lucien Rapp**,** "High Resolution Satellite Imagery and Potential Identification of Individuals", P. Soille, S. Loekken, and S. Albani (Eds.) Proc. of the 2019 conference on Big Data from Space (BiDS'2019), EUR 29660 EN, Publications Office of the European Union, Luxembourg, 2019, p.237-240. https://www.bigdatafromspace2019.org/QuickEventWebsitePortal/2019-conference-on-big-data-from-space-bids19/bids-2019
[76] Regulation (EU) 2016/679 (General Data Protection Regulation) on the protection of natural persons with regard to the processing of personal data and on the free movement of such data, OJ L 119, 04.05.2016
[77] The attribute "identified" refers to a known person, and "identifiable" refers to a person who is not identified yet, but identification is still possible. An individual is *directly* identified or identifiable by reference to "direct or unique identifiers". These "direct and unique identifiers" covers data types which can be easily referenced and associated with an individual, including descriptors such as a name, an identification number or username, location data, card of phone numbers, online identifiers, etc. (art. 4 (1)).
[78] The attribute "identified" refers to a known person, and "identifiable" refers to a person who is not identified yet, but identification is still possible. An individual is *directly* identified or identifiable by reference to "direct or unique identifiers". These "direct and unique identifiers" covers data types which can be easily referenced and associated with an individual, including descriptors such as a name, an identification number or username, location data, card of phone numbers, online identifiers, etc. (art. 4 (1)).

Arguably, a person – as a whole – can be depicted on these pictures, as for the resolution might allow for the identification of a person, considering, for example, the person's height, body type and clothing. Likewise, objects and places (location data) linked to a person could also enable identification of a person via very high-res images (VHR), such as the person's home, cars, boats and others[79]. The lawfulness of its processing needs to be then assured.

As **massive constellations of small satellites**[80] are becoming a staple in LEO, a larger influx of data, observation capabilities and high-quality imagery from EO satellites[81] is expected to become more widely available on a timely basis. EO massive constellations may provide more frequent image capture and updates (capturing a single point several times a day) at a much lower cost. Users can plan both the target and frequency, allowing for a more specific analysis in a particular tracking. Ordinarily, these collected terabytes of data that must be downlinked to a ground station before being processed and reviewed. But now, enabled satellites can carry **mission applications on board, including AI that would conduct that processing on the satellite**[82]. This means that only the most relevant data would be transmitted, not only saving on downlink costs but also allowing ground analysts to focus on the data that matters most. For example, one company has **developed algorithms relying on AI to analyze stacks of images and automatically detect change**, allowing users to track changes to individual properties in any country: *"This machine learning tool, it's constantly looking at the imagery and classifying things it finds within them: that's a road, that's a building, that's a flood, that's an area that's been burned."*[83] Other analytics companies feed visual data into algorithms designed to derive added value from mass images.

AI may be used, in **breach** of EU data protection and other rules, by public authorities or other private entities for mass surveillance. **Very high-resolution (VHR) optical data** may have the same quality as aerial photography, and therefore may **raise respective privacy**[84]**, data protection and ethical issues**.[85]-[86],[87],[88],[89]

In addition, **EO data** may be explored by smart video or **face recognition technologies**[90]-[91] and **combined with other data streams as GPS, security cameras,** etc., thus raising privacy concerns, even if the raw or pre-processed data itself do not.

**Examples of several scenarios can be anticipated where identifiability of individuals can be at stake**: Applying very high-res satellites for scanning the landscape, inspection thereof, capturing images of

---

[79]Aloisio, G. (2017). Privacy and Data Protection Issues of the European Union Copernicus Border Surveillance Service. Master thesis. University of Luxembourg.

[80]The EO constellation will be centered at 600km, which spans a large range of altitudes. It comprises 300 non-maneuverable 3U cubesats so is much smaller in bothtotal areal cross-section and aggregate mass

[81]Popkin, G."Technology and satellite companies open up a world of data", https://www.nature.com/articles/d41586-018-05268-w

[82] http://interactive.satellitetoday.com/via/december-2019/space-2-0-taking-ai-far-out/

[83]http://www.insurancefraud.org/IFNS-detail.htm?key=27644

[84]https://www.technologyreview.com/s/613748/satellites-threaten-privacy/?utm_medium=tr_social&utm_source=twitter&utm_campaign=site_visitor.unpaid.engagement

[85]ITU-T Study Group 17 (SG17), https://www.itu.int/en/ITU-T/about/groups/Pages/sg17.aspx

[86] Von der Dunk, F., "Outer Space Law Principles and Privacy", in Evidence from Earth Observation Satellites: Emerging Legal Issues, Denise Leung and Ray Purdy (editors), Leiden: Brill, pp. 243–258, 2013.

[87] European Space Policy Institute, "Current Legal Issues for Satellite Earth Observation", 2010, p. 38.

[88] Cristiana Santos, Lucien Rapp, "Satellite Imagery, Very High-Resolution and Processing-Intensive Image Analysis: Potential Risks Under the GDPR", *Air and Space Law,* 2019, vol. 44, Issue 3, p. 275–295.

[89] Cristiana Santos, Delphine Miramont, Lucien Rapp**,** "High Resolution Satellite Imagery and Potential Identification of Individuals", P. Soille, S. Loekken, and S. Albani (Eds.) Proc. of the 2019 conference on Big Data from Space (BiDS'2019), EUR 29660 EN, Publications Office of the European Union, Luxembourg, 2019, p.237-240. https://www.bigdatafromspace2019.org/QuickEventWebsitePortal/2019-conference-on-big-data-from-space-bids19/bids-2019

[90] https://www.cnil.fr/sites/default/files/atoms/files/facial-recognition.pdf

[91] See: Facial recognition technology: fundamental rights considerations in the context of law enforcement, https://fra.europa.eu/en/publication/2019/facial-recognition.

buildings, cars, real estate showcasing, stock image production, production of footage for publicity purposes, and suchlike. Those familiar with the area and/or familiar with the individuals who may be in the vicinity may be able to identify them and their movements as well as social patterns. The actual risk prompts by making this data available open-source to be used for **any unforeseen purpose**.

The **European strategy for data**[92] aims a secure and dynamic **data-agile economy** in the world – empowering Europe with data to improve decision-making and better the lives of all its citizens. **The future regulatory framework for AI** in Europe aims to create an '**ecosystem of trust'**. To do so, it must ensure compliance with EU rules, including the rules protecting fundamental rights and consumers' rights, in particular for AI systems that pose a high risk, as explained in this chapter.[93] If a clear European regulatory framework is sought to build trust among consumers and businesses in AI in space, and therefore speed up the uptake of the technology, it is necessary to **be aware of risks of AI in space.**

While AI can do much good, it can also do **harm.** This harm might be both **materia**l (safety and health of individuals, including loss of life, damage to property) and **immaterial** (loss of privacy, limitations to the right of freedom of expression, human dignity, discrimination for instance in access to employment), and can relate to a wide variety of risks. Elaborating upon a forward-looking perspective, harnessing AI technologies in accessing and exploring outer space, as well as engaging in space based commercial activities will, in all likelihood**, span a broad array of intended and unintended consequences** flowing from the use and misuse of such technologies which cannot be downplayed or disregarded. However, it is considered that the two most prominent and complex legal issues are the issue of privacy and data protection on one hand, and liability for erroneous positioning on the other hand[94].

### 3.2.1  Privacy, data protection issues

Employing AI in connection with satellite imaging raises concern relating to personal privacy and data protection. Some of the potential forecasted risks [95] include the following:

- *Ubiquity of "facial recognition data."*[96] Facial recognition data can, potentially, be obtained from a plethora of sources. The facial images collected and registered in a multitude of widely available databases can be used to track movement of people through time and space and therefore constitute a potential source for identifying individuals by an analysis of the images captured by the various facial recognition systems. More generally, any photograph can potentially become a piece of biometric data with more or less straightforward technical processing. The dissemination of data collected by facial recognition devices is taking place amid a context of permanent self-exposure on social media which increases the porosity of facial recognition data. This indicates that a massive amount of data is technically accessible for which AI can potentially be mobilized for facial recognition-based identification

- *Lack of transparency* requires that the data controller inform the data subject of the personal information collected, the purpose of the collection and use of the data. Transparency also entails the imagery operators to inform data subjects of their rights to access, correct and erase the personal data as well as well as the procedure for exercising such rights. The transparency obligation is

---

[92] https://ec.europa.eu/digital-single-market/en/policies/building-european-data-economy
[93] White Paper on Artificial Intelligence - A European approach to excellence and trust (COM(2020) 65 final)
[94] F. von der Dunk, "Legal aspects of navigation - The cases for privacy and liability: An introduction for non-lawyers", Coodinates Magazine, May 2015, http://mycoordinates.org/legal-aspects-of-navigation/
[95] High-Level Expert Group that published Guidelines on trustworthy AI in April 2019[95].
[96] https://www.cnil.fr/sites/default/files/atoms/files/facial-recognition.pdf

- *Data maximization and disproportionality of data processing:* space technology entails the tendency of extensive collection, aggregation and algorithmic analysis of all the available data for various reasons, which hampers the data minimization principle. In addition, irrelevant data is also being collected and archived, undermining the storage limitation principle.

- *Purpose limitation and repurposing of data*. Since data analytics can mine stored data for new insights and find correlations between apparently disparate datasets, space big data is susceptible to reuse for secondary unauthorized purposes, profiling, or surveillance.[97] This undermines the purpose specification principle, which conveys the purpose for which the data is collected must be specified and lawful. As for a repurpose, personal data should not be further processed in a way that the data subject might consider unexpected, inappropriate or otherwise objectionable and, therefore, unconnected to the delivery of the service. Moreover, once the infrastructure is in place, Facial Recognition technology may easily be used for "**function creep**"[98]: which refers to situations when, for instance, the purposes of VHR usage expand, either to additional operations or to additional activities within the originally envisaged operation. It also contemplates when such imagery is disseminated over the internet which naturally increases the risk of the data being reused widely. Given this circumstance it is problematic that the data subject can effectively control the use of the facial recognition data by giving or withholding consent.

- *Retrace.* By analyzing large amounts of data and identifying links among them, AI can be used to **retrace and de-anonymize**[99] data about persons, creating new personal data protection risks even in respect to datasets that per se do not include personal data.

- *Rights of access, correction and erasure*. Results drawn from data analysis may not be representative or accurate if its sources are not subject to proper validation. For instance, AI analysis combining online social media resources are not necessarily representative of the whole population at stake. Moreover, machine learning may contain hidden bias in its programming or software which can lead to inaccurate predictions and inappropriate profiling of persons. Hence, AI interpretation of data collected by high-res images need human validation to ensure the trustworthiness of a given interpretation and avoid the incorrect interpreting an image.. "At best, satellite images are interpretations of conditions on Earth – a "snapshot" derived from algorithms that calculate how the raw data are defined and visualized".[100] This can create a black box, making it difficult to know when or why the algorithm gets it wrong. For example, one recently developed algorithm was designed to identify artillery craters on satellite images – but the algorithm also identified locations that looked similar to craters but were not craters. This demonstrates the need for metrics to assist in in formulating an accurate interpretation of big space data

- *Potential identification of individuals*. For instance, if the footage taken through VHR imaging only shows the top of a person's head and one cannot identify that person without using sophisticated means, it is not personal data. However, if the same image was taken in the backyard of a house with additional imaging analytical algorithms that may enable an identification of the house and/or the owner, that footage would be considered as a personal data. Thus, personal data is very much context-dependent. This scenario escalates with the advances of "ultra-high" definition images being published online, from commercial satellite companies, and the consequential application of big data analytic tools. It might be possible to identify indirectly an individual (and to depict individual households, etc.), when high-res images are combined with other spatial and non-spatial datasets. Thus, while the footage of people may be restricted to "the tops of people's heads", once these images are contextualized by particular landmarks or other information, they may become identifiable. "*The combination of publicly available data pools with high resolution image data, coupled with the integration and analysis capabilities of modern Geographic Information Systems (GIS) disclosing geographic keys such as longitude and latitude, can result in a technological invasion of personal privacy*".[101]

---

[97] https://edps.europa.eu/press-publications/press-news/blog/ai-and-facial-recognition-challenges-and-opportunities_en
[98] https://edps.europa.eu/press-publications/press-news/blog/ai-and-facial-recognition-challenges-and-opportunities_en
[99] White paper on AI
[100] Melinda Laituri, 2018, https://theconversation.com/satellite-imagery-is-revolutionizing-the-world-but-should-we-always-trust-what-we-see-95201

- **Risk to anonymity in the public space**[102]*:* Erosion of anonymity, by public authorities or private organizations, is likely to jeopardize some of the fundamental privacy principles established by the GDRP. Facial recognition in public areas can end up making harmless behavior look suspicious. Wearing a hood, sunglasses or a cap, looking at your telephone or at the ground, can have an impact on the effectiveness of these devices and serve as a basis for suspicion in itself.[103] Additionally, the interface between facial recognition systems and satellite imaging creates the opportunity for an unprecedented level of surveillance, whether by a governmental or private entity. It is not unthinkable that coupling satellite imagery with facial recognition software and other types of technology, such as sound capturing devices, further increases the level of surveillance of people and places.

- **Fallible technologies might produce unfair bias**[104] **and outcomes**[105]-[106]: Like any biometric processing, facial recognition is based on statistical estimates of the match between the elements being compared. It is therefore inherently fallible because it is a probability of match. The French Data Protection explains furthermore that the biometric templates calculated are always different depending on the conditions under which they are calculated (lighting, angle, image quality, resolution of the face, etc.). Every device therefore exhibits variable performance according, on the one hand, to its aims, and, on the other hand, to the conditions for collecting the faces being compared. Space AI embedded devices with facial recognition can thus lead to "false positives" (a person is wrongly identified) and "false negatives" (the system does not recognize a person who ought to be recognized). Depending on the quality and configuration of the device, the rate of false positives and false negatives may vary. The model's result may be incorrect or discriminatory if the training data renders a biased picture reality, or if it has no relevance to the area in question. Such use of personal data would be in contravention of the fairness principle.

- **Transparency and (in)visibility.** This risk applies when individuals on the ground may not know VHR satellites are in operation, and if they do, may be unsure about who is operating them and the purpose for which it is being used, causing somehow discomfort.

---

[101] Chun S., Atluri V., Protecting Privacy from Continuous High-Resolution Satellite Surveillance. In: Thuraisingham B., et al. (eds.) Data and Application Security. IFIP, vol 73. Springer, 2002.
[102] https://www.cnil.fr/sites/default/files/atoms/files/facial-recognition.pdf
[103] https://www.cnil.fr/sites/default/files/atoms/files/facial-recognition.pdf
[104] Joy Buolamwini, Timnit Gebru; Proceedings of the 1st Conference on Fairness, Accountability and Transparency, PMLR 81:77-91, 2018.
[105] https://www.cnil.fr/sites/default/files/atoms/files/facial-recognition.pdf
[106] http://www3.weforum.org/docs/WEF_Framework_for_action_Facial_recognition_2020.pdf

- *Seamless and ubiquitous processing*: VHR combined with facial recognition technologies allows remote, *contactless*[107], data processing. Such "contactless" system, entails that processing devices are excluded from the user's field of vision. It allows the remote processing of data without a person's knowledge, without any interaction nor relationship with the person and therefore without their even being aware of it. Against this scenario, data controllers need to declare the existence of the rights and the procedure for their effective use (Articles 13(2)(b) of the GDPR).

- *Privacy and Non Public Areas:* Using AI with satellite imaging presents issues relating to loss of control over one´s personal information and activities,[108]-[109] which encompasses the right of individuals to move in their own home (yards and gardens) and/or other non public places without being identified, tracked or monitored. [110]

- *Privacy of association:* This refers to the freedom of people to associate with others. [111] It is related also with the fact that footage might indicate, for example, the number of adults living in a household (based on the number of vehicles) or the clues as to their relationships. Satellite imaging and AI provide the opportunity to ascertain and/or monitor personal associations.

- **Verification of compliance.** The specific characteristics of many AI technologies, including opacity ('black box-effect'), complexity, unpredictability and partially autonomous behavior, may make it hard to verify compliance with, and may hamper the effective enforcement of, rules of existing EU law meant to protect fundamental rights. Enforcement authorities and affected persons might lack the means to verify how a given decision made with the involvement of AI in space was taken and, therefore, whether the relevant rules were respected. Individuals and legal entities may face difficulties with effective access to justice in situations where such decisions may negatively affect them.

### 3.2.2 Ethical Issues

- *Discrimination. P*rofiling consists of "pattern recognition, comparable to categorization, generalization and stereotyping."[112] VHR satellite imaging combined with analytic technologies *can* lead to discriminatory profiling.[113] Also, satellite based VHR may be used more in relation to certain populations or areas which are less likely to be able to effectively voice or act upon those concerns (e.g., marginalized populations or areas).With the use of ML and data mining, individuals might be clustered according to generic behaviors, preferences and other characteristics, even without being identified.[114] Profiling ultimately involves creating derived or inferred data, occasionally leading to incorrect and biased decisions (discriminatory, erroneous and unjustified, regarding for instance, their behavior, health, creditworthiness, recruitment, insurance risk, etc.).[115]

---

[107] https://www.cnil.fr/sites/default/files/atoms/files/facial-recognition.pdf
[108] Nissenbaum, H., Privacy in Context: Technology, Policy, and the Integrity of Social Life , (Stanford, CA:Stanford Law Books, 2010), 70-72.
[109] Solove, D."Understanding Privacy", (Cambridge, MA: Harvard University Press, 2008), 24-29.
[110] Finn, R. L., Wright D. et al. "Seven types of Privacy", in Gutwirth, S., Leenes, R., de Hert, P., Poullet, Y. (Eds.), European Data Protection: Coming of Age, Springer, Dordrecht, 2013, p. 16.
[111] Finn, R. L., Wright D. et al. "Seven types of Privacy", in Gutwirth, S., Leenes, R., de Hert, P., Poullet, Y. (Eds.), European Data Protection: Coming of Age, Springer, Dordrecht, 2013, p. 16.
[112] Hildebrandt, M., & Gutwirth, S. *Profiling the European Citizen.* 2008, Springer.
[113] ICO, Big Data, Artificial Intelligence, Machine Learning and Data Protection, UK, 2017
[114] van der Sloot, B., "Privacy in the Post-NSA Era:Time for a Fundamental Revision?" *Journal of Intellectual Property, Information Technology and E-Commerce Law, 5*(1), 2014.
[115] Edwards, L. Privacy, security and data protection in smart cities: a critical EU law perspective. European Data Protection Law Review, Berlin, Vol. 2, 2016, pp. 28-58.

- *Public dissatisfaction*: this refers to the possibility that people could become disillusioned with surveillance and imagery use based on the possibility that they are compromising privacy and data protection rights or that they are feeling "over-run" by such technology.

- *Chilling effect:* The situations where individuals might be unsure about whether they are being observed (even if no VHR satellites are processing data), and they hence attempt to adjust their behavior accordingly.[116]

- *Imbalance*. In a prospective scenario, space technologies might produce situations of imbalance, where data subjects are not aware of the fundamental elements of data processing and related consequences, being unable to negotiate their information, which leads to a side consequence of enhanced information asymmetry. Even exercising the right to be forgotten seems hard to apply. Images captured for use in Street View may contain sensitive information about people who are unaware they are being observed and photographed.[117]

If these risks materialize, the lack of clear requirements and the characteristics of space based AI technologies makes it difficult to trace back potentially problematic decisions made with the involvement of AI systems. This in turn may make it difficult for persons having suffered harm to obtain compensation under the current EU and national liability legislation[118]

# 4 Limitations in the space treaties in determining the law applicable to intelligent systems and services

Limitations naturally exist in the space law treaty regime as it employs broad legal principles accompanied by ambiguous terms and provisions. Moreover, the regime does not sufficiently reflect the metamorphosis on the access, use of outer space being engineered by the escalation and diversification of space activities engaged in by private actors and other non-governmental entities, and the technological advancements such as AI. The lack of international standardization in the space law treaty regime generally surfaces when some sort of fortuitous event occurs such as i) damage to a space asset[1] and ii) an act which increases the hazards of the access and use of outer space.[2] This Section will analyze and discuss the limitations associated with the State centric space legal regime, jurisdiction and choice of substantive law in the restrictive State centric space law treaty regime.

## 4.1 State centric space legal regime

The **space law treaty regime does not impose any direct obligations on non-governmental entities**. Instead, it vests all responsibilities and obligations to only one class of space actor, a State. For instance, the

---

[116] Finn, R. L., Wright D. et al. "Seven types of Privacy", in Gutwirth, S., Leenes, R., de Hert, P., Poullet, Y. (Eds.), European Data Protection: Coming of Age, Springer, Dordrecht, 2013, p. 16.
[117] President's Council of Advisors on Science and Technology, Big Data and Privacy: A Technological Perspective, Report to the President (May 2014),
[118] White Paper On Artificial Intelligence - A European approach to excellence and trust, https://ec.europa.eu/info/publications/white-paper-artificial-intelligence-european-approach-excellence-and-trust_en

[1] Injury or harm is not limited to physical collision with a space object but includes conduct such as jamming a satellite transmission, hijacking a satellite signal, or seizing command and control of a space object.

[2] The creation of space debris by testing an anti-satellite weapon is an example of an act increasing the hazards for the access and use of space.

Outer Space Treaty Article VI establishes that the outer space activities of non-governmental entities are subject to restrain and control by States and not directly by the treaty regime (provided that the non-governmental actor's space activity does not involves piracy, genocide or any other recognized international crime).

The most fundamental limitation is apparent from Outer Space Treaty Article XIII which recognizes that the Treaty provisions apply only to the activities of States which includes international governmental organizations and its relations. Limiting the obligations and remedies associated with space activities to States essentially **relegates the space law treaty regime to the rule of politics rather than the rule of law.** As the space economy matures, a space law legal regime directly applicable to all space actors and rooted in the rule of law, rather than politics, will become a necessity. Until there is a movement beyond a State centric space legal regime, international space law will suffer the limitations associated with restricted direct application as well as limitations with respect to areas such as jurisdiction and the choice of the applicable substantive law.

## 4.2 Jurisdictional limitation

There is no international body with jurisdiction to compel adjudicating space based dispute between or among States and bind the States to a judgment. International jurisdiction over space based disputes rests on the consent of all States that are parties to the matter.[1] Moreover, since the space law treaties do not impose direct obligations and duties on non-governmental entities, there is no basis for international jurisdiction over a non-governmental space actor provided that the non-governmental actor's space activity does not involves piracy, genocide or any other recognized international crime.

The jurisdictional limitation of the space law treaty regime is further apparent whenever a private individual or non-governmental entity desires to directly pursue a remedy for harm flowing from a space activity. Under that circumstance, jurisdiction is determined by State law unless the parties to the dispute consent to private arbitration. Although space law precludes a State from exercising sovereignty in outer space, space law incorporates international law which recognizes a State's power to exercise jurisdiction over extraterritorial acts under certain circumstances.[2] Accordingly, State law governing jurisdiction can arguably extend to AI related disputes arising in space. Even more so, a State can enact specific legislation granting its courts or agencies jurisdiction over AI based disputes arising from space activities. In either event, jurisdiction is proper only if such legislation satisfies one of the five tentacles for extraterritorial jurisdiction.[3]

An example of a State extending such jurisdiction in the space context is seen by the United States enacting a statute which criminalizes any intentional or malicious interference with "the authorized operation of a communications or weather satellite or obstructs or hinders any satellite transmission."[4] Noticeably, the statute does not: 1) limit its application to a satellite launched by or registered to the United States 2) limit is application to affecting United States national security, United States citizens, United States economic interests or any other interest of or nexus to the United States.[5] In the absence of universal jurisdiction over interference with a satellite, the statute's jurisdictional scope appears to be overbroad and extends beyond the jurisdiction reach consistent with international law.

---

[1] Marc S. Firestone, Problems in the Resolution of Disputes Concerning Damage Caused in Outer Space, 59 Tul. L. Rev. 747, 763 (1985)

[2] *United States v. Ali*, 885 F.Supp.2d 17, 25-26 rev. in part on other grounds 885 F.Supp.2d 55 (D.D.C. 2012). Customary international law generally recognizes five tentacles for a State exercising jurisdiction over the extraterritorial conduct of a non-governmental entity. As the United States judiciary has recognized in the context of a piracy case, the five tenets are territorial, protective, national, passive personality, and universal. *Id*.

[3] *See Id.*

[4] 18 U.S.C. 1367

In any event, there is no international harmonization or standardization for jurisdiction over AI related disputes arising from space activities that do not cause damage as defined by the Liability Convention or when a remedy is being directly pursued by a private person or a non-governmental entity.

### 4.3  Limitations of Space Law

The space treaties do not address the use of AI, nor any international treaty, regulates AI in space. This means that domestic legislation must serve as the principal source for the substantive law relating to the use of AI in space. This lack of international regulation of AI poses **potential complex problems relating to the applicable substantive law in disputes involving the use of AI in space.** For instance, if the use of AI or an intelligent space object causes damage to another space object which is cognizable under the Liability Convention, it is unclear what substantive law applies to determine to the issues relevant to resolving the merits of the claim such as the standard of care and what constitutes fault. Liability Convention Article XVIII provides that the "Claims Commission shall decide the merits of the claim for compensation and determine the amount of compensation payable, if any." However, neither Article XVIII nor any other provision of the Liability Convention indicates what substantive law is used to decide the merits and determine the compensation issue. Is the appropriate substantive law the domestic law of: 1) the launching State of the space object causing the damage, 2) the Registry State of the space object causing the damage, 3) the State that owned or whose national owned the damaged space object, 4) the Registry State of the damaged space object, 5) the home State of the software developer for the AI used by the space object that caused the damage, or 6) substantive law formulated by the Claims Commission.[12]

Similar choice of substantive law problems exists if the dispute involves a space-based injury which is not subject to the Liability Convention or if an injured non-governmental entity decides to directly pursue a claim for injury arising from a space activity. If such a claim is brought in the judiciary of State, then that State's conflict of law provisions may serve as a guide for determining which substantive law applies. To this extent, it becomes an open question as to whether the Liability Convention's liability scheme for allocating fault can apply to private persons or non-governmental entities. The judiciary in Belgium and the United States have each adopted customary international law principles embodied in an international treaty as the substantive law for resolving a dispute between two private parties arising in the international arena of the high seas.[13] Thus, the Liability Convention's fault allocation scheme may conceivably be used in proceedings for space-based damage or harm arising from the use of AI. However, that does not eliminate the choice of law dilemma associated with determining causation, fault, and other merit related issues.

As seen, the lack of substantive law at the international level limits the ability of the space law treaty regime to establish a harmonious or uniform legal standard for deciding claims involving AI related space based damage or harm.

Given that the Liability Convention employs fault-based liability for extraterrestrial damage caused by a space object, it is sensible and practical that the same liability scheme should be applicable in a legal action involving extraterrestrial harm attributable to AI which is outside the scope of the Liability Convention or in which a non-governmental entity is a party. Otherwise, there will not be any international standard for, among other things, allocating fault and determining liability for extraterrestrial harm arising from space activities. The lack of international standardization means that the plethora of potential substantive law choices renders the choice of substantive law a critical issue. State laws can vary based on whether a State is a common law jurisdiction like the United States and Great Britain, a civil system as employed by European States, an Islamic law State, a State which practices some form of Marxism or some other political or legal system.

---

[5] George Anthony Long, *Legal Basis for a State's Use of Police Power Against Non-Nationals to Enforce Its National Space Legislation* at 3, 70th IAC, Washington, DC, (Oct. 23, 2019 )

[12] Liability Convention Article XVI(3) allows the Claims Commission to determine its own procedure which should include how it chooses the applicable substantive law.

[13] *Castle John v. NV Mabeco*, 77 ILR 537 (Belgium Court of Cassation 1986); *Institute of Cetacean Research* v. *Sea Shepherd Conversation Society,* 708 F.3d 1099 (9th Cir. 2013). The two cases involved plaintiffs seeking injunctive and declaratory relief against a non-State actor for conduct alleged to constitute piracy under international law.

Selecting from the buffet of substantive law choices in matters involving AI is complicated by the fundamental concept that the space law treaty regime, like all State legal systems, is rooted in controlling and regulating decisions, acts, errors, and omissions of a person or people even if made in the guise of a juridical person.[14] AI is machine conduct. This fundamental distinction between AI conduct and human conduct is currently confronting legal systems of technologically advanced States.

The United States is a technologically advanced State that is struggling with ascertaining the right "fit" for legal actions arising from an event involving or associated with AI. Generally, in the United States legal actions seeking compensation for harm caused by a device or machine allege either negligence of the owner/operator or are based on a theory of products liability.[15] However, either theory necessitates a determination of fault based on human conduct. Negligence necessitates human involvement.[16] Products liability concerns a defect in software design or manufacturing and failures to warn of reasonable foreseeable injury.[17] A design defect occurs when a foreseeable risk of harm exists and the designer could have avoided or reduced the risk by utilizing a reasonable alternative design.[18] A manufacturing design occurs when a product is not produced according to specifications.[19] Failure to warn arises when the responsible party fails to "provide instructions regarding how to safely use the software.[20]

Liability for an AI design defect can be either strict liability or fault liability depending upon the particular industry or application using the AI.[21] However, the EC White Paper on AI seemly adopts the fault based approach for injury caused by an AI system. In suggesting that products liability law may not be a "good fit" for AI related injury, the White Paper recognizes that "it may be difficult to prove that there is a defect in the product, the damage that has occurred and the causal link between the two."[22] It further notes that "there is some uncertainty about how and to what extent the Product Liability Directive applies in the case of certain types of defects, for example if these result from weaknesses in the cybersecurity of the product." [23]

Nevertheless, since strict and fault liability are predicated on human conduct, there is an emerging perspective that perhaps a new liability scheme needs to be adopted for AI. Two new liability concepts for AI are some form of legal personhood for AI and substituting the "reasonable man" standard used in current jurisprudence with "robot common sense."[24] There is also suggestion that agency law be used in connection with AI systems as the autonomous machine is actually an agent of the owner or operator. Regardless of how much a new liability standard may be needed for AI, especially in the context of outer space, any such development, in all likelihood, will have to emerge from the domestic law of States. This reflects a substantial limitation in the space law treaty regime's developing a uniform substantive law governing fault liability for space based damage caused by a space object using AI.

---

[14] See *supra* at ______.

[15] Dr. Iria Giuffrida, *Liability for Ai Decision-Making: Some Legal and Ethical Considerations*, 88 Fordham L. Rev. 439, 443 (2019)

[16] *Supra* at ____.

[17] Jason Chung & Amanda Zink, *Hey Watson - Can I Sue You for Malpractice? Examining the Liability of Artificial Intelligence in Medicine*, 11 Asia-Pacific J. of Health L. Pol'y & Ethics 51, 68 (Nov. 2017), http://eible-journal.org/index.php/APHLE/article/view/84.; Megan Sword, *To Err Is Both Human and Non-Human*, 88 UMKC L. Rev. 211, 224 (2019)

[18] *Id.*

[19] *Id.*

[20] *Id.*

[21] Megan Sword, *supra* Note ___, 88 UMKC L. Rev. at 224.

[22]

[23] Id.

# 5  Elements of legal methodology for determining the law applicable to intelligent systems and services

Legal methodology is a way of reaching a legal result in a coherent and deductive way.[1] The most common legal methodology employs a three prong approach which consists of 1) Method of Description , 2) Method of Conceptual Analysis and 3) Method of Evaluation.[2] This section will examine AI in space in light of these three prongs.

"Method of Description" describes the state of affairs as it exists at present."[3] The previous Section articulates the current status of jurisdiction and choice of law relating to disputes involving the use of AI in space.

**Method of Conceptual Analysis**
This method concerns an abstract idea or theory and usually involves two concepts which are: "'(1) analysis of the existing conceptual framework of and about law; (2) construction of new conceptual frameworks with accompanying terminologies.'"[4] Since the previous section focused on the existing conceptual framework and its limitations, this section will focus on examining a new conceptual framework for determining the applicable jurisdictional and substantive law in space based disputes caused by an AI system, at least when one party to the dispute is an EU member State or entity. This approach would be consistent with the EC White Paper on AI which recognizes the necessity of avoiding fragmented and divergent national rules relating to the use of AI within its markets or by juridical entities of its member States.
Accordingly**,** the ruling delivered by the Court of Justice of the European Union (CJEU) on December 19, 2019, regarding the online accommodation-sharing platform Airbnb[119] could serve as an important benchmark for establishing jurisdictional boundaries. It may provide important tangential clues for extending the existing frames of reference used towards determining the applicable legal regimes governing **AI systems and services in space. Although the CJEU ruling pertains exclusively to terrestrial online platforms providing intermediation consumer related services** - the extent of which falls considerably below the scope of **in-orbit servicing** -, the extrapolation of this judgment to extra-terrestrial matters provides a potential alternative route to determining the legal regime of service providers and the larger issue of territoriality and state jurisdiction in space[120].

---

[24] Dr. Iria Giuffrida, *supra* Note ___, 88 Fordham L. Rev. at 448-450.

[1] Kai Schadbach, *The Benefits of Comparative Law: A Continental European View*, 16 Boston Univ. International Law Journal 331, 367 (1998)

[2] Prof. Miodrag Jovanović, *Legal Methodology & Legal Research and Writing - A Very Short Introduction* at 1-2, available at http://147.91.244.8/prof/materijali/jovmio/mei/Legal%20methodology%20and%20legal%20research%20and%20writing.pdf

[3] *Id* at 2.

[4] *Id* quoting Robert S. Summers, *The New Analytical Jurists,* 41 New York University Law Review 865 (1966).

119  Aff. C-390/18, YA and Airbnb Ireland UC versus Hotelière Turenne SAS et Association pour un hébergement et un tourisme professionnel (AHTOP) et Valhotel, Concl. Maciej Szpunar, Press Release n°162/19.

120  In the legal context of the European Union, this judgment is particularly significant in that it achieves this result without attempting to redefine the distribution of powers such as operated by the Lisbon Treaty, where space activities are concerned. Indeed, space as well as the attendant technological research & development activities remain fully entrenched within the realm and

## 5.1 Overview of the solution adopted by the Court of Justice of the European Union

The CJEU provides a legal characterization of Airbnb's activities in view of the recommendations set forth by the Directive on electronic commerce (E-Commerce Directive), concluding that the Airbnb *platform* fits the definition of "*information society services*", as stated in article 2(a) of the E-Commerce Directive.

In attempting to give a legal characterization to Airbnb's service offering, the Court carries out a comprehensive examination of Airbnb's online marketplace as well as its wider business model. It ultimately finds that the defendant's digital platform provides a service of direct intermediation, which is supplied remotely, via electronic means and linking potential lessors and lessees, by providing them with the opportunity to connect and to facilitate their entry into contractual agreements over future transactions.

Although the Airbnb ruling seems at *prima facie* far removed from the realms of space and **AI systems and services,** it offers, nonetheless, a potentially new framework through which to examine broader issues of jurisdiction, extraterritoriality and the determining of appropriate legal regimes to novel phenomena spawned by emerging technologies that elude regulatory oversight.

In this context, the December 19, 2019 Airbnb ruling by the CJEU is a landmark decision that may prompt new insights into how **jurisdictional conflicts may be litigated in the future.** In particular, the decision raises two distinct avenues for inquiry, namely: i) characterizing the platform as a neutral vehicle (as in delivery system) devoid of inherent liability and unconnected to jurisdiction; and ii) the territorialization of the service that is delivered. In other words, the applicable legal regime to a platform - whether that platform is deployed on earth or in outer space - is pegged to the content (the purveyor and/or beneficiary of the services) rather than the container (the physical platform and its operator).

**(a). Emphasis on the nature of the service provided**

What seems important to the Court is not so much the features or functionality of the platform used as the nature of the service provided. What it regards as constituting an information society service is the purpose of the service: to put potential tenants in contact, in return for payment, with professional or non-professional landlords offering short-term accommodation services so that the former can book accommodation.

The fact that this service is provided by means of an electronic platform seems less important than the fact that it is provided at a distance, by electronic means, or that it is provided at the individual request of its recipients, on the basis of an advertisement disseminated by the lessor and an individual request from the tenant interested in the advertisement.

The platform only intervenes in the Court's reasoning as the technical support of the service whose main characteristic for the Court is to be provided remotely. The Court of Justice of the European Union notes that "*through an electronic platform*" (§47), although it acknowledges that this technical support plays an essential role in the service, noting that "*the parties come into contact* [with each other] *only through the electronic platform of the same name*" (§47).

**(b). Unbundling of intermediation and hosting services**

---

jurisdiction of shared competence between the Union and its members states, in complete agreement with article 4 of the Treaty on the Functioning of the European Union, an area over which "the Union shall have competence to carry out activities and conduct a common policy; however, the exercise of that competence shall not result in Member States being prevented from exercising theirs".

This approach is all the more interesting in that the reasoning of the Court of Justice of the European Union concerning the Airbnb electronic platform articulates a second argument: the intermediation service provided by Airbnb by means of the eponymous electronic platform must be dissociated from the real estate transaction "*in so far as it does not consist solely in the immediate provision of accommodation*" (§54). In the Court's view, it consists more in making available on an electronic platform "*a structured list of short-term accommodation (...) corresponding to the criteria adopted by persons seeking short-term accommodation*", so that that service and hence the platform itself are regarded only as "*an instrument facilitating the conclusion of contracts relating to future transactions*" (§53).

As the Court points out, "*it is the creation of such a list for the benefit of both guests with accommodation for rent and those seeking such accommodation which is the key feature of the electronic platform managed by Airbnb Ireland*" (§53).

Put differently by the Court itself, the service provided by Airbnb Ireland by means of its electronic platform "*cannot be regarded as merely ancillary to an overall service falling within a different legal classification, namely the provision of accommodation*" (§54). Nor is it indispensable to the provision of that accommodation directly provided by the lessors, whether professional or non-professional. It only provides one more channel, in addition to other ways and means for the parties to the accommodation contract to meet and conclude.

By recognising its independence, the Court makes it a service providing additional support which serves the objectives of competition and, consequently, of the market, especially since the electronic platform does not intervene in the determination of the price of the hosting service. It is merely a means of facilitation, including all the associated services (photographs of the asset rented, optional instrument for estimating the rental price in relation to market averages taken from the platform, rating system for lessors and lessees), considered as part of "*the collaborative logic inherent in intermediation platforms, which allows, on the one hand, housing applicants to make a fully informed choice from among the housing offers proposed by landlords on the platform and, on the other hand, allows landlords to be fully informed about the seriousness of the tenants with whom they are likely to engage*" (§60).

**(c). Consecration of the law of the country of establishment of the service provider**

By classifying the intermediation service provided by the Airbnb platform as an information society service for the reasons given above, the Court of Justice of the European Union makes it subject to the aforementioned Directive 2000/31. It means that, "*in order to ensure effectively the freedom to provide services and legal certainty for service providers and their recipients (...)*", "*these information society services must in principle be subject to the legal regime of the Member State in which the service provider is established*" (Recital 22).

The attachment of an activity, whether terrestrial or space-based, to the jurisdiction of a State implies the submission of that activity to the legal system of that State. According to the logic of the Internal Market of the European Union, this connection may be that of "*the State in which the service provider is established*". Indeed, since the legal orders of each Member State are supposed to integrate the provisions of the regulations or directives of the European Parliament and the Council, these legal orders are made up of harmonised legislative or regulatory texts.

This is all the more true since the principle of primacy of Community law gives precedence to the European rule over national law and since this European rule is itself directly applicable. The European citizen can therefore obtain its application by the national court whether or not the national law is in conformity with European law.

This is why, in the logic of European integration, the principle adopted for determining the law applicable to a service activity is that of the law of the country in which the service provider is established or, in the case of broadcasting by means of satellite systems, the law of the country in which the signal is transmitted.

A comparable reasoning can be articulated with regard to a platform deployed in space. We shall even see in the following developments that the application of the law of the country of origin of the service provided by means of an intelligent space platform is preferable, in our view, to the law of the country of consumption of the service provided.

**(d). Obligation of prior notification of national provisions**

The Airbnb decision is also interesting in the reminder of the obligation for Member States to notify their national legislation when it is more restrictive than the European rule. This is an interesting idea that can be transposed into international relations. Moreover, a comparable practice exists in the air transport sector, which leaves States sovereign over their respective airspaces and in the name of this sovereignty allows them within the ICAO to make known differences between their national legislations and international rules. These differences are accepted and respected as such on condition that they have been notified to ICAO. The same mechanism could be transposed to the field of space.

In the Airbnb judgment, the European Court of Justice does not proceed differently. It sets aside the national law of the country of consumption of the service, in this case French law, on two separate grounds: the first arises from the principle of the free movement of information society services between Member States, which the Court of Justice considers to be one of the objectives of Directive 2000/31, going so far as to point out that "this objective is pursued by means of a mechanism for monitoring measures liable to undermine it" (§91); the second is a corollary of the first, since it follows from the obligation imposed on Member States by Directive 2000/31 to notify the Commission of measures restricting or liable to restrict the free movement of Information Society services prior to their entry into force.

The Court points out that this obligation to notify does not merely cover "*a mere information requirement*", but corresponds in fact to "*a procedural requirement of a substantive nature justifying the non-applicability to individuals of non-notified measures restricting the free movement of information society services" (§94). As the Court also points out, this is indeed "a standstill l" obligation on the part of the State intending to adopt a measure restricting the freedom to provide an information society service*" (§93).

In its judgment of 19 December 2019, the Court of Justice of the European Union does not reject this eventuality. Quite the contrary. It recognises that, in the extension of the provisions of Article 3 §4 of Directive 2000/31, Member-States have the option of taking measures derogating from the principle of free movement of information society services with regard to a given information society service falling within the coordinated field. However, in addition to the procedural obligation of notification referred to above, it lays down three substantive conditions on which it intends to exercise control (§84):

- the restrictive measure concerned must be necessary in order to guarantee public policy, the protection of public health, public security or consumer protection, AND,

- it must be taken against an Information Society service which effectively undermines or constitutes a serious and grave risk of undermining these objectives, AND,

- it must be proportionate to these objectives.

**(e). Key Takeaways from the Airbnb Case**

From the standpoint of space law, this ruling is especially significant as its rationale may be extended to space platforms that are assembled in outer space and are used to provide AI systems and services in space that exhibit inconclusive connections to terrestrial jurisdictions.

Such stations and platforms, regardless of their complexity, purpose and functionality, remain, in essence, supporting tools and instruments designed to facilitate the provision of a service. In other words, they are the means to an end. As such, the CJEU retains its role as a deciding factor in helping courts determine the true nature and scope of an information-communication technology-related service, whether on Earth or in outer space. In other words, when examining the concept of platform, the Court excludes all metonymical reasoning and retains the nature of the service that is being provided via that platform, rather than the platform as a vehicle for delivering that service, as the primary consideration in making a legal characterization. Further, this ruling argues that even in such circumstances where the provision of a service must *in fine* be conflated with the medium that is used to deliver it[121], the provision of the service must be considered first, over the medium that is used to deliver it, which must be considered second.

## 5.2 Method of Evaluation

The method of evaluation concerns examining "whether rules work in practice, or whether they are in accordance with desirable moral, political, economic aims, or, in comparative law, whether a certain harmonization proposal could work, taking into account other important divergences in the legal systems concerned."[1] As discussed below, the *Airbnb* case may be applicable to disputes involving AI systems and services in space.

A similar rationale can be equally replicated in the context of AI systems and services in space **supplied** via outer-space platforms powered by AI technologies. Drawing on a concept that has been common to both telecommunication law and electronic communication law ever since the industry's opening to market competition, the support service delivered by a given platform is in and of itself a bearer service (or data service) that brings an infrastructure to a user's disposal. Such a service must be distinguished from the global service provision supplied through the platform-as-a-medium; as the service becomes increasingly dematerialized, it too becomes progressively dissociated from its medium.

In the case of in-orbit servicing, if the bulk of the service is delivered in orbit[122], the foreseeable legal challenge lies in accurately identifying the substance and the nature of this service provision and in defining its governing legal regime. That must apply even where this service provision merges with its delivering platform to such an extent that a platform governed by artificial intelligence becomes materially indivisible from the services that it is designed to deliver. From a legal standpoint, the delivery of such a service calls for a distinct characterization that falls under the authority of the principles governing the activities of states in the exploration and use of outer space, including the moon and other celestial bodies (i.e. the Outer Space Treaty).

This particular requirement raises the larger issue of defining the boundaries of the legal forum and the jurisdictional competence of the state—i.e. the range of applicability of national laws over matters exceeding the traditional purview of national legislation. The CJEU's judgment of December 19, 2019 on the matter of Airbnb's digital platform offers an important contribution in this particular regard as well. In characterizing the intermediation service delivered by the Airbnb digital platform as an information society service, for the reasons discussed previously, the CJEU places the defendant's business under the scope of Directive 2000/31. This directive lays down that "*in order to improve mutual trust between Member States [and] to effectively guarantee freedom to provide services and legal certainty for suppliers and recipients of services, such information society services should in principle be subject to the law of the Member State in which the service provider is established*"[123].

---

[121] Likely because the contracting parties could only establish contact through the intermediation of this service/tool.

[1] Prof. Miodrag Jovanović, *supra* Note ___, at 2

[122] Be it remote computation; temperature-controlled storage; maintenance operations; rescue missions; remote sensing and Earth observation; big data storage; etc

Beyond the basic principles of freedom to provide services and legal certainty[124], binding in-orbit service delivering provided via a smart space platform to the legal jurisdictional authority and to the legal regime of a particular state, provides a fresh outlook on the leading doctrine established by article VIII of the OST[125]. Where the weighing of the various of connecting factors enables an Earth or space-bound activity to be ascribed to a given national jurisdiction, that activity is bound by the legal regime of that state. Therefore, in keeping with the EU internal market rationale, this jurisdictional basis can be established as the "*the state in which the service provider is domiciled*"[126].

Within the particular context of European law, considering that the legal regime of each member state is required to integrate the statutory and regulatory provisions issued by the European Parliament and the Council, all of the national legislative and regulatory frameworks are supposed to be harmonized across European jurisdictions, in compliance with the overarching principle of primacy of Community law[127]. This principle requires that the European rule of law must always prevail over national law where a conflict of laws exists, and that European regulations have direct application within national jurisdictions. As a result, every European citizen has the prerogative to avail him or herself of the right to petition a national court to enforce the application of a European statutory or regulatory provision over national law, regardless of whether the national law is compliant with European legislation or not.

As consistent with the guiding values of European integration, the basic legal principle used in determining the appropriate legal regime applicable to the provision of a commercial service, tethers the legal forum to the service provider's place of establishment, or, in the case of satellite frequency broadcasting, to the law of the country from which the signal is emitted. For our purposes, this latter criterion can be a particularly helpful connecting factor where—in the case of sophisticated ICT semi-autonomous applications developed by international teams cutting across traditional jurisdictions—the original place of establishment cannot be conclusively be determined.

In light of the preceding discussion, one proposed way out of the jurisdictional quandary raised by emerging intelligent technologies in outer space would be to bind the provision of the service to the legislation and to the jurisdictional forum of the beneficiary of that service—i.e. the customer or the consumer of that service. Such an approach has the advantage of bringing an added level of clarity to the determination of the appropriate legal forum and to the lingering difficulty of establishing clear connecting factors that bind orbital operators to terrestrial jurisdictions.

This latter situation arises when the country of registration is designated as the sole applicable jurisdiction where the "customer" is an actual space object that is subject to mandatory registration. This particular situation faces many challenges in the context of the digital economy intersecting with the space industry. Rethinking liability around service users and service purveyors might be a way forward that is more in line with the direction that the industry as a whole seems to be taking.

---

123       Directive 2000/31/CE , Recital n°22.
124       Which are not unique to the internal European Union market, and which apply in equal measure to terrestrial commerce as they do to outer space commerce
125       With its twofold implication jurisdictional boundary and government control
126      Please see Jean-Marie de Poulpiquet de Branscavel in *L'immatriculation des satellites*, PhD thesis, University Toulouse-Capitole, December 2018
127       Declarations annexed to the Final Act of the Intergovernmental Conference which adopted the Treaty of Lisbon, (December, 13, 2007). Declaration 17 concerning primacy. Consolidated protocols, annexes and declarations attached to the treaties of the European Union/Declarations.

This proposed solution might also put a stop to the growing practice where, due to the difficulties of tracking and controlling activities of private operators in space, many states are starting to "*relaxing the registration and supervision of corporations* [incurring risk] *of possible liability*" and failing to the duty of care imposed by the treaties (Zhao, 2004, op. cit.).

Going forward, the applicable legislation could be the national jurisdiction of the natural or legal entity that benefits either directly or indirectly from the service that is being supplied in orbit.

Such a system would usher in an unprecedented level of transparency and legal certainty to all stakeholders involved and would further benefit from the existing legal scholarship and regulatory frameworks already in place in other areas of international and regulation law to streamline oversight mechanisms while stimulating industrial development. With particular regard to state subsidies and EU community state aid rules, for instance, there is ample legal expertise and established jurisprudence to help lawmakers trace international finance networks, to determine state accountability and to expose hollow intermediaries[128].

Finally, such a solution would provide the required flexibility necessary to enable any concerned state to anticipate and to introduce appropriate oversight and control mechanisms with its own legislation, which lines up with the observable trend where states are relying ever more on national legislation rather than international consensus to regulate competitive markets.

# 6    AI techniques to support Space Law

The triad of "law, space, and AI" is an instance of the triad "law, science, and technology". The intersection of these latter three is quite extensive exceeding the limits of this whole handbook. The relation between the first triad and this chapter is fairly similar. There are several open questions and developments in AI reflecting on law and legal compliance. Those reflecting the legal questions raised in this chapter so far are, of course, are relevant to be considered. For instance, the developments of machine learning (federated learning, transfer learning, generative adversarial networks) or the real-time analysis in big data in general are approaches that might facilitate GDPR compliance are relevant in space applications too. Also, the big question of liability related to automated decision-making or machine learning algorithms, or the possible accountability of autonomous agents is the open question of la. This means, for now, answering the question entails reliance on State law and general principles of international law, when space law is inapplicable or uncertain. Space is an area where the future answer is of great significance because of the increasing reliance on AI in connection with space activities. Autonomous space agents will need to reason about legal obligations under the applicable law in formulating decisions. Accordingly, the general questions of legal reasoning and the relevant formal systems and other AI applications should involve the same process discussed in the XYZ chapters of this Handbook. In what follows, we will focus on the applicable approaches introduced in this book on the current and foreseeable state(s) of space law.

## 6.1    Legal knowledge representation in the space AI domain

Modelling a consensually shared domain of space law into a machine-readable way increases its chances for such legal knowledge to be interconnected across the Web and used in different applications. As the normative sources are multiple and heterogeneous in this domain, it makes **ontologies** the amenable way for mapping[129] this body of knowledge. Therefore, the inter-relations and framework of space law makes it a natural area for knowledge representation, sharing and reuse.

---

128    See L.Rapp and Ph.Terneyre, Public Business Law, 2020, Lamy Kluwer, n°774 et seq.

**Legal fragmentation amenable for ontologies**

The space law treaty regime is complemented and supplemented **by national legislation** of each country that provides sufficient clarify and densifies provisions on activities that are not directly addressed in the vague, imprecise, or overly broad formulated and ambiguous provisions of the Outer Space Treaty.[130] An analysis of some of the most significant space legislation immediately confirms that space law does not necessarily embody **a unified single text of legislative value**, but combination of **scattered normative legal texts** (that do not all enclose the same legal value), **i.e., legal fragmentation.** And while we can observe some intention of keeping some questions subjected to international treaties and others referred completely to national jurisdictions, as we saw above, in the legal practice, these questions are highly interconnected requiring thoroughfare between the supposedly separate areas.

**Convergence of legal mechanisms as classes of a space law ontology**

Alongside the heterogeneity of forms and contents of national legal texts—reflecting at the same time the legal traditions of each state, their degree of involvement in the space economy and, more and more often, their willingness to differentiate themselves from other nations by offering more favorable conditions for space traders (forum shopping)[11]—their **convergence** is reflected in the most relevant[131] legal mechanisms that are regulated in each different jurisdiction. The following eight provision-types represent the basic schema (domain-specific classes) for any space law ontology that describes the knowledge embedded in the different legal documents.

1. authorization and licensing
2. continuous supervision of non-governmental activity
3. liability
4. insurance
5. space debris removal
6. state strategic interest
7. registration process or registry
8. transfer of ownership in space

**6.2   Relevant Knowledge resources**

To develop an ontology, the elicitation of the most authoritative relevant[132] knowledge resources in this specific legal domain is required. We have hence elicited both non-ontological and ontology-based resources to be further engineered.

**Non-ontological resources.** Non-ontological resources (NOR) in the legal domain are knowledge resources whose semantics have not been formalized yet by means of an ontology but have related semantics which allows interpreting the legal knowledge they hold[133]. In fact, using non-ontological resources from the space domain that convey consensus portray benefits, e.g. interoperability in terms of the vocabulary used, information browse/search, decrease of the knowledge acquisition bottleneck, reuse, among others. The following non-exhaustive domain-expert resources are recommended for they comprise highly reliable domain-related content from websites of organizations embodying domain knowledge. They serve as domain knowledge resources for the development of a space ontology. They are structured and semantically rich taxonomies that serves to annotate the data elements in an ontology. Reuse of these resources can enable a common terminology development and harmonization, a high-level taxonomy for space legal concepts and terms, and thus means to characterize space law. In concrete, we suggest that these resources can enable automatic ontology population.

---

[129] Humphreys L., Santos C., Di Caro L., Boella G., van der Torre L., Robaldo L. (2015), Mapping Recitals to Normative Provisions in EU Legislation to Assist Legal Interpretation, in Frontiers in Artificial Intelligence and Applications (FAIA, 2015), Vol. 279, Legal Knowledge and Information Systems, A. Rotolo (Ed.), IOS Press, p. 41-49, DOI 10.3233/9781-61499-609-5-41
[130] https://www.thespacereview.com/article/3523/1
[131]
[132] V. Opijnen M., Santos C. (2017), On the Concept of Relevance in Legal Information Retrieval, Artificial Intelligence and Law Journal, DOI: 10.1007/s10506-017-9195-8, https://link.springer.com/article/10.1007/s10506-017-9195-8

- Space Legal Tech[134], a tool representing the regulations and national space agencies of 100 countries (as depicted in figure 1)
- NASA Thesaurus (accessible in machine readable form)[135] and Taxonomy[136] documenting a high level set of terms that can be used for mapping together varying data structures
- UCS Satellite Database[137]
- Glossaries, as the ESA Science Glossary[138]
- ESA Earth Observation Knowledge Navigational System[139], a knowledge management system for EO imagery
- Catalog of Space Law Terms[140]
- "Spationary", a work in progress database with structured business and space law concepts and terms (as illustrated in figure 2)

---

[133] Santos C., Casanovas P., Rodríguez-Doncel V., Van der Torre L. (2016), Reuse and Reengineering of Non-ontological Resources in the Legal Domain, AI Approaches to the Complexity of Legal Systems (AICOL) International Workshops 2015–2017, Revised Selected Papers, Springer

[134] http://spacelegaltech.chaire-sirius.eu/

[135] https://www.sti.nasa.gov/nasa-thesaurus/#.XpOFTogzY2w

[136] United States. (2009) *NASA Taxonomy 2.0*. United States. [Archived Web Site] Retrieved from the Library of Congress, https://www.loc.gov/item/lcwaN0014329/.

[137] In-depth details on the 2,218 satellites currently orbiting Earth, including their country of origin, purpose, and other operational details, https://www.ucsusa.org/nuclear-weapons/space-weapons/satellite-database#.W586g85KjtQ

[138] https://solarsystem.nasa.gov/basics/glossary/, https://www.space4water.org/glossary, https://www.factmonster.com/math-science/space/universe/space-glossary, http://www.telespazio.com/glossary-glossario, https://www.esa.int/Our_Activities/Space_Transportation/Glossary, https://amazing-space.stsci.edu/glossary/, https://solarsystem.nasa.gov/basics/glossary/

[139] M. Zingler, R. di Marcantonio, Navigating Through Earth Observation Knowledge, ESA Bulletin 96, Nov 1998. http://www.esa.int/esapub/bulletin/bullet96/ZINGLER.pdf

[140] https://swfound.org/media/99172/guide_to_space_law_terms.pdf

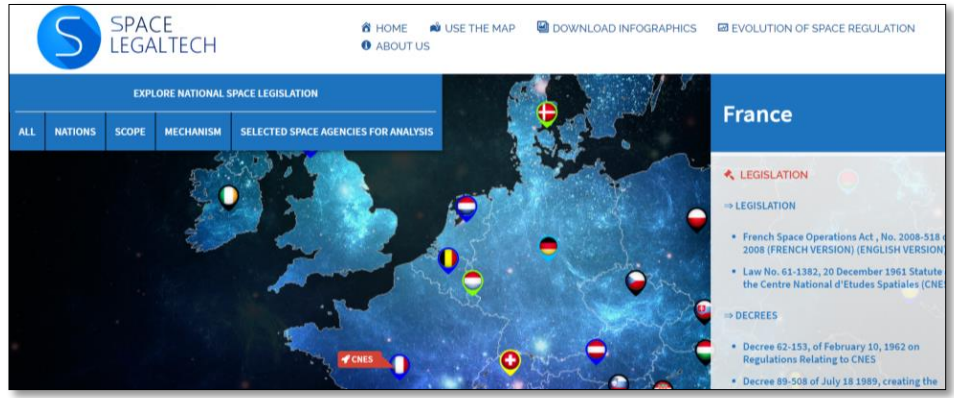

Figure 1 National space legislation from France[141]

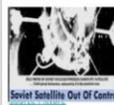

Figure 2 Excerpt of the concept of liability from the Spationary

**Ontological Resources.** We leverage from existing space-related ontological resources (semantically structured information in this domain) can be re-used or extended to any ontology modeling space law, which means that classes and/or instantiations from existing ontologies can be imported.

| Ontological resources | Description |
|---|---|
| Ontology for space object[142] | analysis of the category of Space Object and subcategories. Space objects include artificial objects such a spacecraft, space stations; and natural space objects |
| Ontology-based knowledge management for space data[143] | discusses aspects of ontological engineering in knowledge management architectures for space data |
| Ontology for Satellite Databases[144] | Offers a domain-specific terminology and knowledge model for space data systems. Where data is drawn from multiple sensors or databases, ontologies should foster information fusion via this backbone terminology |

---

[141] Screenshot from the Space LegalTech, http://spacelegaltech.chaire-sirius.eu/use-the-map/
[142] Robert Rovetto, Space Object Ontology, https://philarchive.org/archive/ROVSOO
[143] Rovetto, Robert J, Ontology-based knowledge management for space data, Computer [0018-9162] Rovetto, R J yr:2017 vol:35 iss:11 pg:56
[144] Rovetto, Robert J. "An Ontology for Satellite Databases." Earth Science Informatics 10.4 (2017): 417–427

| Space Surveillance Ontology in XML Schema[145] | capture data structure, content, and semantics in a targeted military domain of space surveillance |
| --- | --- |
| Orbital Debris Ontology[146] | seeks to support orbital debris remediation by ontologically modeling orbital debris, developing accurate and reusable debris classification |
| Space Situational Awareness Ontology[147] | domain coverage of all space objects in the orbital space environment together with relevant Space Situational Awareness (SSA) entities |
| NASA Sweet Ontologies[148] | set of approximately 200 modular ontologies, collectively consisting of approximately 6000 category terms intended to provide a knowledge base to represent Earth science data and knowledge |

In furtherance of any ontological artifact, members of the space community are to provide domain expertise, including verifying the accuracy of the knowledge expressed by the logical formalizations in the ontologies.

**Ontology-based resources in the legal domain**

Building on the assessed risks in the privacy and data protection domain in Section 3.2, we aim to reuse the ontologies which model the concepts concerning the protection of personal data and which are already explained in detail in this book, like the Data Protection Ontology[149], PrivOnto[150], PrOnto[151], GDPRtEXT (GDPR text extensions)[152]. Other core legal domains ontologies are also of use, such as Eurovoc, Legal RuleML[153], ELI.

# 7 Conclusion

In this chapter we have discussed how "intelligent" systems and services raise legal problems, starting with the law applicable thereto, privacy and data protection and liability issues. These legal challenges call for solutions which the international treaties in force are not sufficient to determine and implement. For this reason, a legal methodology is suggested that makes it possible to link intelligent systems and services to a system of rules applicable to them. We also propose legal informatic tools amenable to be applied to space law.

---

[145] A Space Surveillance Ontology Captured in an XML Schema", October 2000, Mary K. Pulvermacher, Daniel L. Brandsma, John R. Wilson, MITRE, Center for Air Force C2 Systems, Bedford, Massachusetts

[146] R.J. Rovetto, An Ontological Architecture for Orbital Debris Data, Earth Science Informatics, 9(1) (2015) 67-82, Springer

[147] Rovetto, Robert J., Kelso, T.S. Preliminaries of a Space Situational Awareness Ontology. (2016) 26th AIAA/AAS Space Flight Mechanics meeting, Napa, California, Feb 17 2016.

[148] https://github.com/ESIPFed/sweet

[149] Bartolini C., Muthuri R., Santos C. (2015), Using ontologies to model data protection requirements in workflows, Proc. of the 9th Int. Work. on Juris-informatics (JURISIN, 2015), Otake M., Kurahashi S., Ota Y., Satoh K., Bekki D. (eds) New Frontiers in Artificial Intelligence. JSAI-isAI 2015. Lecture Notes in Computer Science, vol 10091, p 27-40, Springer

[150] Alessandro Oltramari, Dhivya Piraviperumal, Florian Schaub, Shomir Wilson, Sushain Cherivirala, Thomas B Norton, N Cameron Russell, Peter Story, Joel Reidenberg, and Norman Sadeh. Privonto: A semantic framework for the analysis of privacy policies. Semantic Web, (Preprint):1–19, 2018.

[151] Monica Palmirani, Michele Martoni, Arianna Rossi, Cesare Bartolini, and Livio Robaldo. Pronto: Privacy ontology for legal reasoning. In International Conference on Electronic

[152] Harshvardhan J Pandit, Kaniz Fatema, Declan OSullivan, and Dave Lewis. Gdprtext-gdpr as a linked data resource. In European Semantic Web Conference, pages 481–495. Springer, 2018

[153] Monica Palmirani, Guido Governatori, Antonino Rotolo, Said Tabet, Harold Boley, and Adrian Paschke. Legalruleml: Xml-based rules and norms. In Rule-Based Modeling and Computing on the Semantic Web, pages 298–312. Springer, 2011.